\def\temp{1.35}%
\let\tempp=\relax
\expandafter\ifx\csname psboxversion\endcsname\relax
  \message{PSBOX(\temp)}%
\else
    \ifdim\temp cm>\psboxversion cm
      \message{PSBOX(\temp)}%
    \else
      \message{PSBOX(\psboxversion) is already loaded: I won't load
        PSBOX(\temp)!}%
      \let\temp=\psboxversion
      \let\tempp=\endinput
    \fi
\fi
\tempp
\message{by Jean Orloff: loading ...}
\let\psboxversion=\temp
\catcode`\@=11
%
%
\def\psfortextures{
\def\PSspeci@l##1##2{%
\special{illustration ##1\space scaled ##2}%
}}%
\def\psfordvitops{
\def\PSspeci@l##1##2{%
\special{dvitops: import ##1\space \the\drawingwd \the\drawinght}%
}}%
\def\psfordvips{
\def\PSspeci@l##1##2{%
\d@my=0.1bp \d@mx=\drawingwd \divide\d@mx by\d@my
\includegraphics{##1\space}}}%
\def\psforoztex{
\def\PSspeci@l##1##2{%
\special{##1 \space
      ##2 1000 div dup scale
      \number-\psllx\space\space \number-\pslly\space\space translate
}}}%
\def\psfordvitps{
\def\dvitpsLiter@ldim##1{\dimen0=##1\relax
\special{dvitps: Literal "\number\dimen0\space"}}%
\def\PSspeci@l##1##2{%
\at(0bp;\drawinght){%
\special{dvitps: Include0 "psfig.psr"}
\dvitpsLiter@ldim{\drawingwd}%
\dvitpsLiter@ldim{\drawinght}%
\dvitpsLiter@ldim{\psllx bp}%
\dvitpsLiter@ldim{\pslly bp}%
\dvitpsLiter@ldim{\psurx bp}%
\dvitpsLiter@ldim{\psury bp}%
\special{dvitps: Literal "startTexFig"}%
\special{dvitps: Include1 "##1"}%
\special{dvitps: Literal "endTexFig"}%
}}}%
\def\psfordvialw{
\def\PSspeci@l##1##2{
\special{language "PostScript",
position = "bottom left",
literal "  \psllx\space \pslly\space translate
  ##2 1000 div dup scale
  -\psllx\space -\pslly\space translate",
include "##1"}
}}%
\def\psforptips{
\def\PSspeci@l##1##2{{
\d@mx=\psurx bp
\advance \d@mx by -\psllx bp
\divide \d@mx by 1000\multiply\d@mx by \xscale
\incm{\d@mx}
\let\tmpx\dimincm
\d@my=\psury bp
\advance \d@my by -\pslly bp
\divide \d@my by 1000\multiply\d@my by \xscale
\incm{\d@my}
\let\tmpy\dimincm
\d@mx=-\psllx bp
\divide \d@mx by 1000\multiply\d@mx by \xscale
\d@my=-\pslly bp
\divide \d@my by 1000\multiply\d@my by \xscale
\at(\d@mx;\d@my){\special{ps:##1 x=\tmpx cm, y=\tmpy cm}}
}}}%
\def\psonlyboxes{
\def\PSspeci@l##1##2{%
\at(0cm;0cm){\boxit{\vbox to\drawinght
  {\vss\hbox to\drawingwd{\at(0cm;0cm){\hbox{({\tt##1})}}\hss}}}}
}}%
\def\psloc@lerr#1{%
\let\savedPSspeci@l=\PSspeci@l%
\def\PSspeci@l##1##2{%
\at(0cm;0cm){\boxit{\vbox to\drawinght
  {\vss\hbox to\drawingwd{\at(0cm;0cm){\hbox{({\tt##1}) #1}}\hss}}}}
\let\PSspeci@l=\savedPSspeci@l
}}%
%
%
\newread\pst@mpin
\newdimen\drawinght\newdimen\drawingwd
\newdimen\psxoffset\newdimen\psyoffset
\newbox\drawingBox
\newcount\xscale \newcount\yscale \newdimen\pscm\pscm=1cm
\newdimen\d@mx \newdimen\d@my
\newdimen\pswdincr \newdimen\pshtincr
\let\ps@nnotation=\relax
{\catcode`\|=0 |catcode`|\=12 |catcode`|
|catcode`#=12 |catcode`*=14
|xdef|backslashother{\}*
|xdef|percentother{
|xdef|tildeother{~}*
|xdef|sharpother{#}*
}%
\def\R@moveMeaningHeader#1:->{}%
\def\uncatcode#1{%
\edef#1{\expandafter\R@moveMeaningHeader\meaning#1}}%
\def\execute#1{#1}
\def\psm@keother#1{\catcode`#112\relax}
\def\executeinspecs#1{%
\execute{\begingroup\let\do\psm@keother\dospecials\catcode`\^^M=9#1\endgroup}}%
\def\@mpty{}%
\def\matchexpin#1#2{
  \fi%
  \edef\tmpb{{#2}}%
  \expandafter\makem@tchtmp\tmpb%
  \edef\tmpa{#1}\edef\tmpb{#2}%
  \expandafter\expandafter\expandafter\m@tchtmp\expandafter\tmpa\tmpb\endm@tch%
  \if\match%
}%
\def\matchin#1#2{%
  \fi%
  \makem@tchtmp{#2}%
  \m@tchtmp#1#2\endm@tch%
  \if\match%
}%
\def\makem@tchtmp#1{\def\m@tchtmp##1#1##2\endm@tch{%
  \def\tmpa{##1}\def\tmpb{##2}\let\m@tchtmp=\relax%
  \ifx\tmpb\@mpty\def\match{YN}%
  \else\def\match{YY}\fi%
}}%
\def\incm#1{{\psxoffset=1cm\d@my=#1
 \d@mx=\d@my
  \divide\d@mx by \psxoffset
  \xdef\dimincm{\number\d@mx.}
  \advance\d@my by -\number\d@mx cm
  \multiply\d@my by 100
 \d@mx=\d@my
  \divide\d@mx by \psxoffset
  \edef\dimincm{\dimincm\number\d@mx}
  \advance\d@my by -\number\d@mx cm
  \multiply\d@my by 100
 \d@mx=\d@my
  \divide\d@mx by \psxoffset
  \xdef\dimincm{\dimincm\number\d@mx}
}}%
%
\newif\ifNotB@undingBox
\newhelp\PShelp{Proceed: you'll have a 5cm square blank box instead of
your graphics.}%
\def\s@tsize#1 #2 #3 #4\@ndsize{
  \def\psllx{#1}\def\pslly{#2}%
  \def\psurx{#3}\def\psury{#4}
  \ifx\psurx\@mpty\NotB@undingBoxtrue
  \else
    \drawinght=#4bp\advance\drawinght by-#2bp
    \drawingwd=#3bp\advance\drawingwd by-#1bp
  \fi
  }%
\def\sc@nBBline#1:#2\@ndBBline{\edef\p@rameter{#1}\edef\v@lue{#2}}%
\def\g@bblefirstblank#1#2:{\ifx#1 \else#1\fi#2}%
{\catcode`\%=12
\xdef\B@undingBox{
\def\ReadPSize#1{
 \readfilename#1\relax
 \let\PSfilename=\lastreadfilename
 \openin\pst@mpin=#1\relax
 \ifeof\pst@mpin \errhelp=\PShelp
   \errmessage{I haven't found your postscript file (\PSfilename)}%
   \psloc@lerr{was not found}%
   \s@tsize 0 0 142 142\@ndsize
   \closein\pst@mpin
 \else
   \if\matchexpin{\GlobalInputList}{, \lastreadfilename}%
   \else\xdef\GlobalInputList{\GlobalInputList, \lastreadfilename}%
     \immediate\write\psbj@inaux{\lastreadfilename,}%
   \fi%
   \loop
     \executeinspecs{\catcode`\ =10\global\read\pst@mpin to\n@xtline}%
     \ifeof\pst@mpin
       \errhelp=\PShelp
       \errmessage{(\PSfilename) is not an Encapsulated PostScript File:
           I could not find any \B@undingBox: line.}%
       \edef\v@lue{0 0 142 142:}%
       \psloc@lerr{is not an EPSFile}%
       \NotB@undingBoxfalse
     \else
       \expandafter\sc@nBBline\n@xtline:\@ndBBline
       \ifx\p@rameter\B@undingBox\NotB@undingBoxfalse
         \edef\t@mp{%
           \expandafter\g@bblefirstblank\v@lue\space\space\space}%
         \expandafter\s@tsize\t@mp\@ndsize
       \else\NotB@undingBoxtrue
       \fi
     \fi
   \ifNotB@undingBox\repeat
   \closein\pst@mpin
 \fi
\message{#1}%
}%
%
%
\def\psboxto(#1;#2)#3{\vbox{%
   \ReadPSize{#3}%
   \advance\pswdincr by \drawingwd
   \advance\pshtincr by \drawinght
   \divide\pswdincr by 1000
   \divide\pshtincr by 1000
   \d@mx=#1
   \ifdim\d@mx=0pt\xscale=1000
         \else \xscale=\d@mx \divide \xscale by \pswdincr\fi
   \d@my=#2
   \ifdim\d@my=0pt\yscale=1000
         \else \yscale=\d@my \divide \yscale by \pshtincr\fi
   \ifnum\yscale=1000
         \else\ifnum\xscale=1000\xscale=\yscale
                    \else\ifnum\yscale<\xscale\xscale=\yscale\fi
              \fi
   \fi
   \divide\drawingwd by1000 \multiply\drawingwd by\xscale
   \divide\drawinght by1000 \multiply\drawinght by\xscale
   \divide\psxoffset by1000 \multiply\psxoffset by\xscale
   \divide\psyoffset by1000 \multiply\psyoffset by\xscale
   \global\divide\pscm by 1000
   \global\multiply\pscm by\xscale
   \multiply\pswdincr by\xscale \multiply\pshtincr by\xscale
   \ifdim\d@mx=0pt\d@mx=\pswdincr\fi
   \ifdim\d@my=0pt\d@my=\pshtincr\fi
   \message{scaled \the\xscale}%
 \hbox to\d@mx{\hss\vbox to\d@my{\vss
   \global\setbox\drawingBox=\hbox to 0pt{\kern\psxoffset\vbox to 0pt{%
      \kern-\psyoffset
      \PSspeci@l{\PSfilename}{\the\xscale}%
      \vss}\hss\ps@nnotation}%
   \global\wd\drawingBox=\the\pswdincr
   \global\ht\drawingBox=\the\pshtincr
   \global\drawingwd=\pswdincr
   \global\drawinght=\pshtincr
   \baselineskip=0pt
   \copy\drawingBox
 \vss}\hss}%
  \global\psxoffset=0pt
  \global\psyoffset=0pt
  \global\pswdincr=0pt
  \global\pshtincr=0pt 
  \global\pscm=1cm 
}}%
%
%
\def\psboxscaled#1#2{\vbox{%
  \ReadPSize{#2}%
  \xscale=#1
  \message{scaled \the\xscale}%
  \divide\pswdincr by 1000 \multiply\pswdincr by \xscale
  \divide\pshtincr by 1000 \multiply\pshtincr by \xscale
  \divide\psxoffset by1000 \multiply\psxoffset by\xscale
  \divide\psyoffset by1000 \multiply\psyoffset by\xscale
  \divide\drawingwd by1000 \multiply\drawingwd by\xscale
  \divide\drawinght by1000 \multiply\drawinght by\xscale
  \global\divide\pscm by 1000
  \global\multiply\pscm by\xscale
  \global\setbox\drawingBox=\hbox to 0pt{\kern\psxoffset\vbox to 0pt{%
     \kern-\psyoffset
     \PSspeci@l{\PSfilename}{\the\xscale}%
     \vss}\hss\ps@nnotation}%
  \advance\pswdincr by \drawingwd
  \advance\pshtincr by \drawinght
  \global\wd\drawingBox=\the\pswdincr
  \global\ht\drawingBox=\the\pshtincr
  \global\drawingwd=\pswdincr
  \global\drawinght=\pshtincr
  \baselineskip=0pt
  \copy\drawingBox
  \global\psxoffset=0pt
  \global\psyoffset=0pt
  \global\pswdincr=0pt
  \global\pshtincr=0pt 
  \global\pscm=1cm
}}%
%
\def\psbox#1{\psboxscaled{1000}{#1}}%
\newif\ifn@teof\n@teoftrue
\newif\ifc@ntrolline
\newif\ifmatch
\newread\j@insplitin
\newwrite\j@insplitout
\newwrite\psbj@inaux
\immediate\openout\psbj@inaux=psbjoin.aux
\immediate\write\psbj@inaux{\string\joinfiles}%
\immediate\write\psbj@inaux{\jobname,}%
%
%
\def\toother#1{\ifcat\relax#1\else\expandafter%
  \toother@ux\meaning#1\endtoother@ux\fi}%
\def\toother@ux#1 #2#3\endtoother@ux{\def\tmp{#3}%
  \ifx\tmp\@mpty\def\tmp{#2}\let\next=\relax%
  \else\def\next{\toother@ux#2#3\endtoother@ux}\fi%
\next}%
%
%
\let\readfilenamehook=\relax
\def\re@d{\expandafter\re@daux}
\def\re@daux{\futurelet\nextchar\stopre@dtest}%
\def\re@dnext{\xdef\lastreadfilename{\lastreadfilename\nextchar}%
  \afterassignment\re@d\let\nextchar}%
\def\stopre@d{\egroup\readfilenamehook}%
\def\stopre@dtest{%
  \ifcat\nextchar\relax\let\nextread\stopre@d
  \else
    \ifcat\nextchar\space\def\nextread{%
      \afterassignment\stopre@d\chardef\nextchar=`}%
    \else\let\nextread=\re@dnext
      \toother\nextchar
      \edef\nextchar{\tmp}%
    \fi
  \fi\nextread}%
\def\readfilename{\bgroup%
  \let\\=\backslashother \let\%=\percentother \let\~=\tildeother
  \let\#=\sharpother \xdef\lastreadfilename{}%
  \re@d}%
%
%
\xdef\GlobalInputList{\jobname}%
\def\psnewinput{%
  \def\readfilenamehook{
    \if\matchexpin{\GlobalInputList}{, \lastreadfilename}%
    \else\xdef\GlobalInputList{\GlobalInputList, \lastreadfilename}%
      \immediate\write\psbj@inaux{\lastreadfilename,}%
    \fi%
    \let\readfilenamehook=\relax%
    \ps@ldinput\lastreadfilename\relax%
  }\readfilename%
}%
\expandafter\ifx\csname @@input\endcsname\relax    
  \immediate\let\ps@ldinput=\input\def\input{\psnewinput}%
\else
  \immediate\let\ps@ldinput=\@@input
  \def\@@input{\psnewinput}%
\fi%
\def\nowarnopenout{%
 \def\warnopenout##1##2{%
   \readfilename##2\relax
   \message{\lastreadfilename}%
   \immediate\openout##1=\lastreadfilename\relax}}%
\def\warnopenout#1#2{%
 \readfilename#2\relax
 \def\t@mp{TrashMe,psbjoin.aux,psbjoint.tex,}\uncatcode\t@mp
 \if\matchexpin{\t@mp}{\lastreadfilename,}%
 \else
   \immediate\openin\pst@mpin=\lastreadfilename\relax
   \ifeof\pst@mpin
     \else
     \edef\tmp{{If the content of this file is precious to you, this
is your last chance to abort (ie press x or e) and rename it before
retexing (\jobname). If you're sure there's no file
(\lastreadfilename) in the directory of (\jobname), then go on: I'm
simply worried because you have another (\lastreadfilename) in some
directory I'm looking in for inputs...}}%
     \errhelp=\tmp
     \errmessage{I may be about to replace your file named \lastreadfilename}%
   \fi
   \immediate\closein\pst@mpin
 \fi
 \message{\lastreadfilename}%
 \immediate\openout#1=\lastreadfilename\relax}%
{\catcode`\%=12\catcode`\*=14
\gdef\splitfile#1{*
 \readfilename#1\relax
 \immediate\openin\j@insplitin=\lastreadfilename\relax
 \ifeof\j@insplitin
   \message{! I couldn't find and split \lastreadfilename!}*
 \else
   \immediate\openout\j@insplitout=TrashMe
   \message{< Splitting \lastreadfilename\space into}*
   \loop
     \ifeof\j@insplitin
       \immediate\closein\j@insplitin\n@teoffalse
     \else
       \n@teoftrue
       \executeinspecs{\global\read\j@insplitin to\spl@tinline\expandafter
         \ch@ckbeginnewfile\spl@tinline
       \ifc@ntrolline
       \else
         \toks0=\expandafter{\spl@tinline}*
         \immediate\write\j@insplitout{\the\toks0}*
       \fi
     \fi
   \ifn@teof\repeat
   \immediate\closeout\j@insplitout
 \fi\message{>}*
}*
\gdef\ch@ckbeginnewfile#1
 \def\t@mp{#1}*
 \ifx\@mpty\t@mp
   \def\t@mp{#3}*
   \ifx\@mpty\t@mp
     \global\c@ntrollinefalse
   \else
     \immediate\closeout\j@insplitout
     \warnopenout\j@insplitout{#2}*
     \global\c@ntrollinetrue
   \fi
 \else
   \global\c@ntrollinefalse
 \fi}*
\gdef\joinfiles#1\into#2{*
 \message{< Joining following files into}*
 \warnopenout\j@insplitout{#2}*
 \message{:}*
 {*
 \edef\w@##1{\immediate\write\j@insplitout{##1}}*
\w@{
\w@{
\w@{
\w@{
\w@{
\w@{
\w@{
\w@{
\w@{
\w@{
\w@{\string\input\space psbox.tex}*
\w@{\string\splitfile{\string\jobname}}*
\w@{\string\let\string\autojoin=\string\relax}*
}*
 \expandafter\tre@tfilelist#1, \endtre@t
 \immediate\closeout\j@insplitout
 \message{>}*
}*
\gdef\tre@tfilelist#1, #2\endtre@t{*
 \readfilename#1\relax
 \ifx\@mpty\lastreadfilename
 \else
   \immediate\openin\j@insplitin=\lastreadfilename\relax
   \ifeof\j@insplitin
     \errmessage{I couldn't find file \lastreadfilename}*
   \else
     \message{\lastreadfilename}*
     \immediate\write\j@insplitout{
     \executeinspecs{\global\read\j@insplitin to\oldj@ininline}*
     \loop
       \ifeof\j@insplitin\immediate\closein\j@insplitin\n@teoffalse
       \else\n@teoftrue
         \executeinspecs{\global\read\j@insplitin to\j@ininline}*
         \toks0=\expandafter{\oldj@ininline}*
         \let\oldj@ininline=\j@ininline
         \immediate\write\j@insplitout{\the\toks0}*
       \fi
     \ifn@teof
     \repeat
   \immediate\closein\j@insplitin
   \fi
   \tre@tfilelist#2, \endtre@t
 \fi}*
}%
\def\autojoin{%
 \immediate\write\psbj@inaux{\string\into{psbjoint.tex}}%
 \immediate\closeout\psbj@inaux
 \expandafter\joinfiles\GlobalInputList\into{psbjoint.tex}%
}%
%
%
%
\def\centinsert#1{\midinsert\line{\hss#1\hss}\endinsert}%
\def\psannotate#1#2{\vbox{%
  \def\ps@nnotation{#2\global\let\ps@nnotation=\relax}#1}}%
\def\pscaption#1#2{\vbox{%
   \setbox\drawingBox=#1
   \copy\drawingBox
   \vskip\baselineskip
   \vbox{\hsize=\wd\drawingBox\setbox0=\hbox{#2}%
     \ifdim\wd0>\hsize
       \noindent\unhbox0\tolerance=5000
    \else\centerline{\box0}%
    \fi
}}}%
%
\def\at(#1;#2)#3{\setbox0=\hbox{#3}\ht0=0pt\dp0=0pt
  \rlap{\kern#1\vbox to0pt{\kern-#2\box0\vss}}}%
%
\newdimen\gridht \newdimen\gridwd
\def\gridfill(#1;#2){%
  \setbox0=\hbox to 1\pscm
  {\vrule height1\pscm width.4pt\leaders\hrule\hfill}%
  \gridht=#1
  \divide\gridht by \ht0
  \multiply\gridht by \ht0
  \gridwd=#2
  \divide\gridwd by \wd0
  \multiply\gridwd by \wd0
  \advance \gridwd by \wd0
  \vbox to \gridht{\leaders\hbox to\gridwd{\leaders\box0\hfill}\vfill}}%
%
\def\fillinggrid{\at(0cm;0cm){\vbox{%
  \gridfill(\drawinght;\drawingwd)}}}%
%
%
\def\textleftof#1:{%
  \setbox1=#1
  \setbox0=\vbox\bgroup
    \advance\hsize by -\wd1 \advance\hsize by -2em}%
\def\textrightof#1:{%
  \setbox0=#1
  \setbox1=\vbox\bgroup
    \advance\hsize by -\wd0 \advance\hsize by -2em}%
\def\endtext{%
  \egroup
  \hbox to \hsize{\valign{\vfil##\vfil\cr%
\box0\cr%
\noalign{\hss}\box1\cr}}}%
%
\def\frameit#1#2#3{\hbox{\vrule width#1\vbox{%
  \hrule height#1\vskip#2\hbox{\hskip#2\vbox{#3}\hskip#2}%
        \vskip#2\hrule height#1}\vrule width#1}}%
\def\boxit#1{\frameit{0.4pt}{0pt}{#1}}%
\catcode`\@=12 
%
\psfordvips   

\input harvmac.tex


\font\eightsl=cmsl8
\font\ninerm=cmr9
\font\ninesl=cmsl9
\font\ninebf=cmbx9
\font\twelverm=cmr12
\hbadness=10000
\def\header#1{\headline{\centerline{\eightsl #1}}}
\def\sect#1{\vskip 1cm \centerline{\bf #1} }
\def\subsect#1{\vskip 0.6cm {\bf\noindent #1}}

\def\pn{\par\noindent}
\def\display#1{\vskip1mm \par\indent  #1  \vskip1mm\noindent}
\def\subskip{\vskip 0.3cm}
\def\before{\vskip1pt\line{\hrulefill}}
\def\after{\line{\hrulefill}\vskip1pt}
\def\boxeqn#1{\vcenter{\vbox{\hrule\hbox{\vrule\kern3pt\vbox{\kern3pt
	\hbox{${\displaystyle #1}$}\kern3pt}\kern3pt\vrule}\hrule}}}
\outer\def\bdis{\obeylines\startdisplay}
{\obeylines\gdef\startdisplay#1
  {\catcode`\^^M=5$$#1\halign\bgroup\indent##\hfil&&\qquad##\hfil\cr}}
\outer\def\edis{\crcr\egroup$$}
\def\tablerule{\noalign{\smallskip\hrule\smallskip}}
\def\tabletoprule{\noalign{\hrule\smallskip}}
\def\tablebottomrule{\noalign{\smallskip\hrule}}
\def\Null#1{}

\def\eqnorm#1{\eqno({\rm #1})}
\def\puteqno#1{\hfill\break #1}
\newcount\eqnumber 
\def\theEquationNumber{%
	\the\eqnumber}
\def\num{%
	\global\advance\eqnumber by 1
	\theEquationNumber}
\def\numadd{%
	\theEquationNumber}	

\def\ga{\alpha}
\def\gb{\beta}
\def\gc{\gamma}
\def\gd{\delta}
\def\gep{\epsilon}
\def\gD{\Delta}
\def\gi{\iota}
\def\gk{\kappa}
\def\gl{\lambda}
\def\gL{\Lambda}
\def\go{\omega}
\def\gO{\Omega}
\def\gs{\sigma}
\def\gt{\tau}
\def\gT{\Theta}
\def\gth{\theta}

\def\bc{{\bf c}}
\def\bC{{\bf C}}
\def\bZ{{\bf Z}}
\def\bone{{\bf 1}}
\def\rmd{{\rm d}}
\def\sh{{\rm sh}}
\def\ch{{\rm ch}}
\def\End{{\rm End}}
\def\Id{{\rm Id}}
\def\End{{\rm End}}
\def\Log{{\rm log}}
\def\arg{{\rm arg}}
\def\Hom{{\rm Hom}}
\def\mod{{\rm mod}}
\def\res{{\rm res}}
\def\rank{{\rm rank}}
\def\Tr{{\rm Tr}}
\def\Ker{{\rm Ker}}
\def\Im{{\rm Im}}
\def\for{{\rm for\ }}
\def\Proof{{\it Proof. }}
\def\qed{\hfill {\it q.e.d.}\break}

\def\dy{{\rm d}y}
\def\dx{{\rm d}x}
\def\dz{{\rm d}z}
\def\half{{1 \over 2}}
\def\quar{{1 \over 4}}
\def\sltwo{s\ell(2)}
\def\uqghat{U_q(\widehat g)}
\def\sln{s\ell(n)}
\def\ot{\otimes}
\def\Vtn{{V^{\ot n}}}
\def\Vtq{{V^{\ot q}}}
\def\Vtp{{V^{\ot p}}}
\def\Rc{\check R}
\def\cX{\check X}
\def\cW{\check W}
\def\dcn{g^\vee }
\def\cL{{\cal L}}
\def\lra{\longrightarrow}
\def\da{\downarrow}
\def\uq{U_{q}}
\def\til{\tilde}
\def\qd{\Delta(u)}
\def\tensor{\otimes}
\def\YX{Y(X_r)}
\def\am{^{(a)}_m}
\def\om#1{\Lambda_{#1}}
\def\pr#1{P_{\Lambda_{#1}}}
\def\V#1{V_{\Lambda_{#1}}}
\def\W#1{W^{#1}_1(u)}
\def\Wnu#1{W^{(#1)}_1}
\def\tensor{\otimes}
\def\Rc{\check{R}}
\def\bC{{\bf C}}
\def\ba#1{\overline{#1}}
\def\Id{{ Id}}
\def\inv{^{-1}}

\def\dripol{Drinfel'd polynomial\ }
\def\hwv{highest weight vector}

\def\Dp{Drinfel'd polynomial}
\def\Dps{Drinfel'd polynomials}
\def\xm{x^-}
\def\xp{x^+}
\def\xpm{x^\pm}
\def\xmp{x^\mp}
\def\giso{\ {\buildrel g \over \simeq}\ }
\def\Pbar{\bar P}

\def\rDri{1}
\def\rKRS{2}
\def\rKS{3}
\def\rResii{4}
\def\rKRii{5}
\def\rSk{6}
\def\rBKKW{7}
\def\rZZ{8}
\def\rOW{9}
\def\rOgi{10}
\def\rORW{11}
\def\rLus{12}
\def\rDEM{13}
\def\rBer{14}
\def\rLS{15}
\def\rHH{16}
\def\rBP{17}
\def\rTar{18}
\def\rCheiii{19}
\def\rDrii{20}
\def\rCPi{21}
\def\rCPiii{22}
\def\rMac{23}
\def\rKNS{24}
\def\rBel{25}
\def\rMacii{26}
\def\rZGB{27}
\def\rDGZ{28}
\def\rMP{29}
\def\rFH{30}
\def\rZam{31}
\def\rFZ{32}
\def\rBCDS{33}
\def\rFF{34}
\def\rDGZa{35}
\def\rDori{36}
\def\rDorii{37}
\def\reference{
\bigskip
\pn
{\bf References}	\vskip5pt\ninerm \par\baselineskip=11pt
\item{[\rDri]}{V.G.\ Drinfel'd, Sov.\ Math.\ Dokl.\  {\ninebf 32} (1985) 254;
in {\ninesl Proceedings of the International Congress of Mathematicians,
Berkeley},
(American Mathematical Society, Providence, 1987)}
\item{[\rKRS]}{P.P.\ Kulish, N.Yu.\ Reshetikhin and E.\ K.\ Sklyanin,
Lett.\ Math.\ Phys.\  {\ninebf 5} (1981) 393}
\item{[\rKS]}{P.P.\ Kulish and E.K.\ Sklyanin, J.\ Sov.\ Math.\  {\ninebf
19} (1982) 1596; in
{ Lecture Notes in Physics} {\ninebf 151} (Springer, Berlin, 1982) }
\item{[\rResii]}{Yu.N.\ Reshetikhin, Lett.\ Math.\ Phys.\ {\ninebf 7}
(1983) 205}
\item{[\rKRii]}{A.N.\ Kirillov and N.Yu.\ Reshetikhin,
Zap.\ Nauch.\ Semin.\ LOMI {\ninebf 160} (1987) 211
[J.\ Sov.\ Math.\ {\ninebf 52} (1990) 3156]}
\item{[\rSk]}{E.K.\ Sklyanin,  hep-th/9211111}
\item{[\rBKKW]}{B.\ Berg, M.\ Karowski, V.\ Kurak and P.\ Weiz,
Nucl.\ Phys.\  {\ninebf B134} (1978) 125}
\item{[\rZZ]}{A.B.\ Zamolodchikov and Al.B.\ Zamolodchikov,
Ann.\ Phys.\ {\ninebf 120} (1979) 253}
\item{[\rOW]}{E.I.\ Ogievetsky and P.B.\ Wiegmann,
Phys.\ Lett.\ {\ninebf B168} (1986) 360}
\item{[\rOgi]}{E.I.\ Ogievetsky, J.\ Phys.\ G {\ninebf 12} (1986) L105}
\item{[\rORW]}{E.I.\ Ogievetsky, N.Yu.\  Reshetikhin and P.B.\ Wiegmann,
Nucl.\ Phys.\ {\ninebf B280} [FS 18] (1987) 45}
\item{[\rLus]}{M.\ L\"uscher, Nucl.\ Phys.\ {\ninebf B135} (1978) 1}
\item{[\rDEM]}{H.\ de Vega, H.\ Eichenherr, J.M.\ Maillet,
Nucl.\ Phys.\ {\ninebf B240} [FS12] 377}
\item{[\rBer]}{D.\ Bernard, Commun.\ Math.\ Phys.\ {\ninebf 137} (1991) 191}
\item{[\rLS]}{A.\ LeClair and F.\  Simirnov,
Int.\ J.\ Mod.\ Phys.\ {\ninebf A7} (1992) 2997}
\item{[\rHH]}{F.D.\ Haldane, Z.N.C.\ Ha,
J.C.\ Talstra, D.\ Bernard and V.\ Pasquier, Phys.\ Rev.\ Lett.\
{\ninebf 69} (1992) 2021}
\item{[\rBP]}{D.\ Belnard, M.\ Gaudin, F.D.M.\ Haldane
and V.\ Pasquier, J.\ Phys.\ A:Math.\ Gen.\ {\ninebf 26}
(1993) 5219}
\item{[\rTar]}{V.O.\ Tarasov, Theore.\ Math.\ Phys.\  {\ninebf 61}
 (1984) 163; {\ninebf 63} (1985) 175}
\item{[\rCheiii]}{I.\ Cherednik, in Proc.\ of the XVII International
Conference
on Differential Geometric Methods in Theoretical Physics, Chester,
ed.\ A.I.\ Solomon, (World Scientific, Singapore, 1989)
}
\item{[\rDrii]}{V.G.\ Drinfel'd, Soviet Math.\ Dokl.\  {\ninebf 36} (1988) 212}
\item{[\rCPi]}{V.\  Chari and A.\ Pressley, L'Enseignement Math.\
{\ninebf 36} (1990) 267}
\item{[\rCPiii]}{V.\  Chari and A.\ Pressley, J.\ reine angew.\ Math.\
{\ninebf 417}
(1991)  87}
\item{[\rMac]}{N.J.\ MacKay, J.\ Phys.\ A: Math.\ Gen.\ {\ninebf 25} (1992)
L1343}
\item{[\rKNS]}{A.\ Kuniba, T.\ Nakanishi, J.\ Suzuki, hep-th/9309137,
to appear in Int.\ J.\ Mod.\ Phys.\ {A}}
\item{[\rBel]}{A.A.\ Belavin, Phys.\ Lett.\ {\ninebf B283} (1992) 67}
\item{[\rMacii]}{N.J.\ MacKay, J.\ Phys.\ A: Math.\ Gen.\ {\ninebf 24} (1991)
4017}
\item{[\rZGB]}{R.B.\ Zhang, M.D.\ Gould and A.J.\ Bracken,
Nucl.\ Phys.\ {\ninebf B354} (1991) 625}
\item{[\rDGZ]}{G.W.\ Delius, M.D.\ Gould and Y.-Z.\ Zhang, hep-th/9405030}
\item{[\rMP]}{W.G.\ McKay and J.\ Patera, {\ninesl Tables of Dimensions,
Indices, and Branching Rules for Representations of Simple
Lie Algebras}, (Marcel Dekker, New York and Basel, 1981)}
\item{[\rFH]}{W.\ Fulton and J.\ Harris, {\ninesl Representation
Theory}, (Springer-Verlag, New York, 1991)}
\item{[\rZam]}{A.B.\ Zamolodchikov, Int.\ J.\ Mod.\ Phys.\
{\ninebf A4} (1989) 4235}
\item{[\rFZ]}{V.A.\ Fateev and A.B.\ Zamolodchikov,
Int.\ J.\ Mod.\ Phys.\ {\ninebf A5} (1990) 1025}
\item{[\rBCDS]}{H.W.\ Braden, E.\ Corrigan, P.E.\ Dorey,
and R.\ Sasaki, Nucl.\ Phys.\ {\ninebf B338}
(1990) 689; {\ninebf B356} (1991) 469}
\item{[\rFF]}{B.L.\ Feigin and E.\ Frenkel,
hep-th/9310022 and references therein}
\item{[\rDGZa]}{G.W.\ Delius, M.T.\ Grisaru,
and D.\ Zanon,  Nucl.\ Phys.\ {\ninebf B382}
(1992) 365}
\item{[\rDori]}{P.\ Dorey,
 Phys.\ Lett.\ {\ninebf B312}
(1993) 291}
\item{[\rDorii]}{P.\ Dorey,
 Nucl.\ Phys.\ {\ninebf B358}
(1991) 654; Nucl.\ Phys.\ {\ninebf B374}
(1992) 741
}
}

\Title{HUTP-94/A010, hep-th/9405200}{}
\centerline{\bf Fusion, mass, and representation theory
of the Yangian algebra}
\bigskip\bigskip
\centerline{Tomoki Nakanishi\footnote{$^\dagger$}
{On leave from Department of Mathematics,
Nagoya University, Nagoya, Japan 464}}
\bigskip
\centerline{nakanisi@string.harvard.edu}
\centerline{Lyman Laboratory of Physics}
\centerline{Harvard University}\centerline{Cambridge, MA 02138}


\vskip .3in
Based on the formulation of Drinfel'd, Chari, and Pressley, a technique to
analyze the
structure of  tensor products of the  Yangian algebra representations
is presented. We then apply the results to the $S$-matrix theory of
the  $G\otimes G$-invariant nonlinear $\sigma$-model ($G$-principal chiral
model)
by Ogievetsky, Reshetikhin, and Wiegmann.
We show how the physical data such as mass formula, fusion angle, and
the spins of integrals of motion can be extracted from the Yangian
highest weight representations.

\Date{05/94}

\newsec{Introduction}

The Yangian $Y(g)$ [\rDri] is a Hopf algebra associated to a simple Lie algebra
$g$.

The  structure of the Yangian   first appeared in the study of the spectrum of
the
transfer matrices of a series of integrable spin chains [\rKRS-\rKRii]
in the algebraic Bethe ansatz method
(see [\rSk] for a concise introduction),  in the study of the
elastic S-matrices of  two-dimensional relativistic field theory
[\rBKKW--\rORW],
and in the system with  non-local conserved currents
[\rLus--\rLS].
 These works already exhibited a rich structure and  characteristic properties
  of the
Yangian representations. Several representation
theoretical
ideas, such as $R$-matrix, fusion procedure,
were already  in  [\rKRS].
More recently the Yangian also arose in a long-range interacting
spin chain [\rHH,\rBP].

At the same time the representation theory of  the Yangian
has been developed [\rTar-\rCPiii]. In particular the notion of
highest weight representations,  parallel to those of simple Lie algebras,
was studied  in [\rDrii--\rCPiii].

It may therefore be appropriate and  useful to describe
some of the results in the above mentioned works
by using  highest weight representations. In this paper
we focus on the $S$-matrix theory of
the  $G\tensor G$-invariant nonlinear
$\gs$-model ($G$-principal chiral model) in
[\rOW,\rORW]. It turns out that the highest weight
representations are more than a useful language.
They are directly related to the basis of the $S$-matrix
theory. Our approach is close to [\rBer] in spirit, but with
more emphasis on  $Y(g)$ representation theoretical
aspects.
Some of the issues studied in this paper have been also
addressed in [\rMac].

In section 2 we prepare basic facts on Yangian
highest weight representations.
In section 3 we present a technique to study  tensor products
 of the fundamental representations, including
some new results.
This provides the proof of the fusion contents of the
fundamental representations used in [\rKNS].
In section 4
 we extend the fundamental representations under the extension
of $Y(g)$ by the two-dimensional Poincar\'e algebra [\rBer].
Applications to an $S$-matrix model are given.
First we reproduce  the mass spectrum of
the  $G\tensor G$-invariant nonlinear
$\gs$-model in [\rOW] from the fusion data of $Y(g)$.
 This is a generalization of
 the idea of [\rBel] for $Y(A_n)$ to all $Y(g)$.
 We then show the triangle relation of the fusing angles,
 and prove that all the bound state poles are indeed in the
 physical strip.
 Finally the selection rule for the spins of integrals of motion is
 derived from the consistency of the fusion data.

\newsec{Yangian}
\eqnumber=0

Let $g$ be a simple Lie algebra with an invariant bilinear form $(\ ,\ )$,
and let $\{ I_p\}$ be an orthonormal basis of $g$.
Then, the Yangian $Y(g)$ is the Hopf algebra generated by
elements $x$, $J(x)$, $(x\in g)$ with the relations
$$
\eqalignno{
&J(ax+by) = aJ(x)+bJ(y), \quad a,b\in \bC, 	\qquad
	[x,J(y)]=J([x,y])	,&(2.\num{\rm a})\cr
&	[J(x),J([y,z])]+[J(y),J([z,x])]+[J(z),J([x,y])]
&(2.\numadd{\rm b})	\cr
& 	=\sum_{p,q,r} ([x,I_p],[[y,I_q],[z,I_r]])\{ I_p,I_q,I_r \}	,
\cr
&	[[J(x),J(y)],[z,J(w)]]+[[J(z),J(w)],[x,J(y)]]	 &(2.\numadd{\rm c})\cr
&	= \sum_{p,q,r}\bigl( ([x,I_p],[[y,I_q],[[z,w],I_r]])
	+([z,I_p],[[w,I_q],[[x,y],I_r]])\bigr)\{ I_p,I_q,J(I_r) \}	,\cr
}
$$
where $\{ x_1,x_2, x_3\} = {1\over24}\sum_{\gs}
x_{\gs(1)}x_{\gs(2)}x_{\gs(3)}$ and the sum is over all
the permutations $\gs$ of $\{ 1,2,3 \}$.

The algebra $Y(g)$ is also realized as an algebra  generated  by elements
$\xpm_{ik}, h_{ik}$,  $ (i=1, \dots, r=\rank\, g, k\in \bZ_{\geq0})$ with the
relations
$$
\displaylines{
[h_{ik},h_{jl}]=0,\quad [h_{i0},x^{\pm}_{jl}]=\pm (\ga_i,\ga_j)x_{jl}^\pm,
\quad [x_{ik}^+,x_{jl}^-]=\delta_{ij}h_{ik+l}, \cr
[h_{ik+1},x_{jl}^\pm]-[h_{ik},x_{jl+1}^\pm]=
\pm \half (\ga_i,\ga_j)(h_{ik} x_{jl}^\pm +x^\pm_{jl}h_{ik}), \cr
\hfill [x_{ik+1}^\pm,x_{jl}^\pm]-[x_{ik}^\pm,x_{jl+1}^\pm]=
\pm \half (\ga_i,\ga_j)(x_{ik}^\pm x_{jl}^\pm +x^\pm_{jl}x_{ik}^\pm),
\puteqno{(2.\num)} \cr
\sum_{\gs} [x_{ik_{\gs(1)}}^\pm, [x_{ik_{\gs(2)}}^\pm,
\dots,[x_{ik_{\gs(1-A_{ij})}}^\pm,x_{jl}^\pm]\dots]]=0
\quad {\rm for}\quad i \neq j.\cr
}
$$
where $A_{ij}$ is the Cartan matrix of $g$. See [\rDri,\rDrii--\rCPiii] for
details.
The correspondence from the first to the second realization is given by
$$
\eqalignno{
h_{i}& \mapsto h_{i0}	, \quad \xpm_i\mapsto\xpm_{i0}, \cr
J(h_i) &\mapsto h_{i1} + \quar \sum_{\ga >0} (\ga, \ga_i)
(\xp_{\ga0}\xm_{\ga0} + \xm_{\ga0} \xp_{\ga0})
-\half h_{i0}^2		, & (2.\num) \cr
J(\xpm_i)&\mapsto \xpm_{i1}	\pm \quar
\sum_{\ga >0} \left( [\xpm_{i0},\xpm_{\ga0} ]\xmp_{\ga0}
+ \xmp_{\ga0} [ \xpm_{i0},\xpm_{\ga0}]
	\right)	-\quar (\xpm_{i0} h_{i0} + h_{i0} \xpm_{i0} ),
\cr
}
$$
where $\xpm_{\ga0}$ are the images of  root vectors $\xpm_\ga \in g$,
 $(\xp_\ga,\xm_\ga)=1$ of weights $\pm \ga$
 through the above correspondence.

The advantage of the first realization is that the comultiplication becomes
simple, i.e., for $x\in g$,
$$
\gD(x) = x\tensor 1 + 1 \tensor x	,\quad
\gD(J(x))= J(x) \tensor  1+ 1 \tensor J(x) + \half [x\tensor 1 , \gO ]	,
\eqno(2.\num)
$$
where $\gO=\sum_p I_p\tensor I_p $.
$\gO$ is related to the second
  Casimir element of $g$, $\gO_2=\sum_p I_p I_p$, as
$$
  \gO = \half (\gD (\gO_2) - \gO_2 \tensor 1 - 1\tensor\gO_2 ).
  \eqno(2.\num)
$$
The advantage of the second
one is that it admits the Poincar\'e-Birkhoff-Witt decomposition
$Y(g)= Y^- H Y^+$, where $Y^\pm$ and $H$ are subalgebras
 generated by  $\xpm_{ik}$
and $h_{ik}$. Then it is possible to define the notion of a highest weight
representation  with a highest weight vector $v$ satisfying
$$
	\xp_{ik}\cdot v=0,\quad h_{ik}\cdot v = d_{ik} v, \quad
	d_{ik}\in \bC	.
\eqno(2.\num)
$$

Let $\gt_i ={ (\ga_i,\ga_i) / 2}$ for each simple root $\ga_i$.
There is a fundamental theorem about highest weight representations.

\vtop{
\proclaim Theorem [\rDrii]. 1). Every irreducible finite dimensional
 representation of\/
$Y(g)$ is highest weight. \hfill\break
2).
An irreducible highest weight representation with highest weight
$\{ d_{ik}\}$ is  finite dimensional if
and only if there exist monic polynomials $P_i(u) $  ($i=1, \dots, r$)
such that
$$
{P_i(u+\gt_i)	\over P_i(u)}=1+\sum_{k=0}^\infty d_{ik} u^{-k-1},
\eqno(2.\num)
$$
\par
}
We call  $P_i(u)$'s  the {\it \Dps}. Then
$d_{i0}= \gt_i \deg P_i$
 is the $g$-highest weight  of a $Y(g)$-\hwv. Thus  \Dps\  are a
 natural, but a quite non-trivial generalization of the notion of
  an  integral weight of $g$.

 In this paper we focus on the irreducible representations such that
 $\deg P_i = \delta_{ia}$ for some $a$. Such representations
 are called  the {\it fundamental representations},  and studied
 in [\rCPiii].
 From now on we choose the normalization of the inner product
 such that $(\ga,\ga)=2$ for any long root $\ga$.
 Then the second Casimir of the adjoint representation
 is equal to $2 \dcn$,
 where $\dcn$ is the dual coxeter number of $g$.
 Let $W_a(b)$ ($b\in \bC$) denote the fundamental
 representation with  the \Dps
 $$
 P_a(u)=u-b-{1\over 2}\gt_i +\quar \dcn,
 \qquad P_i(u)=1\quad {\rm for\ } i\neq a,
 \eqno(2.\num)
 $$
 and let $v_a$ be a \hwv\ of $W_a(b)$. Then we have
 $$
 \eqalignno{
 h_{ik}\cdot v_a &= (b+\half\gt_i -\quar \dcn)^k h_{i0}\cdot v_a,
 &	(2.\num{\rm a})\cr
 \xm_{ik} \cdot v_a &=  (b+\half\gt_i -\quar \dcn)^k
 \xm_{i0} \cdot v_a	.
 &	(2.\numadd{\rm b})\cr
 }
 $$
The first equality is a direct consequence of the definition. The second one
follows from the fact that
 $(\xm_{ak} - (b+ \gt_a/2-\dcn/4)^k \xm_{a0}) \cdot v_a$ is a
  singular vector due to
 (2.9a).

Define the action of $x$ and $J(x)$ on $W_a(b)$ by the correspondence (2.3).
Then  from (2.9a) and (2.9b)  we get (c.f.\ Lemma 3.4 of [\rCPiii])
 $$
  J(x)\cdot v_a = b x \cdot v_a
\eqno(2.\num)
 $$
for all $x =\xpm_i,h_i$.
To derive this, we have used the
 formula,
$$
\sum_{\ga>0} (\ga,\ga_i) h_\ga = \dcn h_i ,	\quad
\sum_{\ga>0} [\xp_\ga, [ \xm_i,\xm_\ga]]=
( (\ga_i,\ga_i)-\dcn )\xm_i	.
\eqno(2.\num)
$$
A simplification of the action of $J(x)$ like (2.10)
does not occur for  a given vector in $W_a(b)$ in general
except for $Y(A_r)$ [\rDri].
Nevertheless
 the action  like (2.10) is true
 in some cases: For instance,
 if there is
  a  vector $v \in W_a(b)$
  and $x \in \{ \xpm_i, h_i\}$ such that
  $x\cdot v = \xm_{i_1}\cdots \xm_{i_n} \cdot v_a\neq 0$ but
$\xm_{i_{\gs(1)}}\cdots \xm_{i_{\gs(n)}}\cdot v_a=0$ for
any  permutation $\gs\neq 1$, then
$J(x)\cdot v = bx\cdot v$ due to
(2.1a) and (2.10).
We use this remark extensively in the next section.

With the above information
 we can study the structure of the tensor products among
the fundamental representations to some extent.
 Eq.\ (2.10) is  the key formula,
because it gives a bridge
between the two realizations of $Y(g)$
in the analysis of highest weight representations.

\newsec{Tensor product and fusion procedure}
\eqnumber=0

\topinsert
\centerline{\psboxscaled{700}{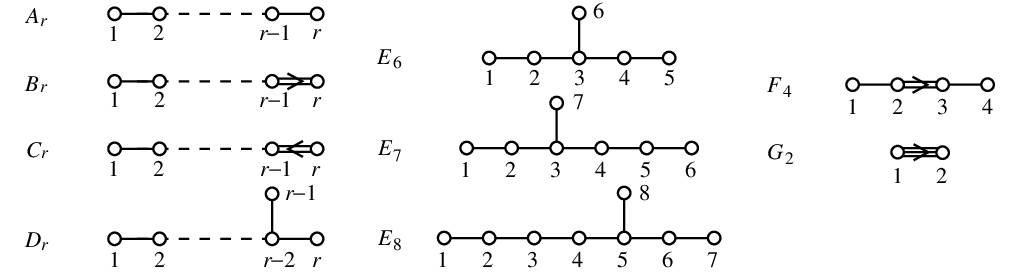}}
{\ninerm
 \centerline{ Figure 1.\quad The  numbering of the simple roots
 of the simple Lie algebras.}
}
\endinsert

In this section we present a way to study
structures of tensor products of
fundamental representations.
We write  the irreducible $g$-module with the highest weight $\gL$
 as
 $V_{\gL}$,
 and
the $i$-th fundamental weight as $\gL_i$  ($\gL_0 = 0$).
 We fix the  numbering of the simple roots as in Fig.1.

\subsec{Multiplicity free case}
We first consider the case when the decomposition of a
tensor product as a $g$-module is multiplicity free
(i.e., the dimension of the
space of the $g$-highest weight vectors of a given weight is at most one).
To make our argument concrete, we use an example in $g=F_4$.
As a
$g$-module
the $Y(F_4)$-module $W_4(b)$  is isomorphic to $V_{\gL_4}$,
 and $J(x)$  acts
on $W_4(b)$ as $b x$ for  any $x\in g$ [\rDri].

Below we shall determine the $Y(g)$-module structure of the tensor product
$$
W_4(b') \tensor W_4(b).
\eqno(3.\num)
$$
We do this in two steps: Step 1).  Find all the $Y(g)$-highest weight
vectors. Step 2). If there is a $Y(g)$-\hwv,
determines the Drinfel'd polynomials of the submodule generated by it.

{\it Step 1).} Since any $Y(g)$-highest weight vector is $g$-highest, we
start by  decomposing $W_4(b')\tensor W_4(b)$ as a $g$-module to
$$
V_{\gL_4} \oplus V_{\gL_3} \oplus V_{2\gL_4} \oplus V_{\gL_1} \oplus V_{\gL_0}.
\eqno(3.\num)
$$
Let $u_4$ and $v_4$ be  $Y(g)$-\hwv s of $W_4(b')$ and
$W_4(b)$.
An explicit form of
a $g$-highest weight vector $w_a$ of each component in (3.2) is given in Table
1.
It is trivial that $w_{2\cdot 4}$ is $Y(g)$-highest.
Our main question is whether the other $g$-highest weight vectors are
$Y(g)$-highest or not. Since the decomposition (3.2) is multiplicity free,
we only
need to check the condition
$$
J(\xp_i)\cdot w_a = 0.
\eqno(3.\num)
$$
For each $\gL_a$ $(a=0,1,3,4)$, we assign
 the weight $\imath(\gL_a)$ such that $(\xp_i \tensor 1) \cdot
w_a \in V_{\imath (\gL_a)}
$ for all $i$. They are  summarized in the following diagram:
$$
\gL_4  \buildrel\imath\over\longrightarrow \gL_3
 \buildrel\imath\over\longrightarrow 2\gL_{4}
 \buildrel\imath\over\longleftarrow \gL_1  \buildrel\imath\over\longleftarrow
\gL_0
\eqno(3.\num)
$$
This graph is essentially the {\it tensor product graph}
in [\rMac,\rMacii--\rDGZ], a convenient way to keep track
of the submodule structure.
Because
$J(x)$ acts  by $bx$ on {\it any} vector in $W_4(b)$, we have
$$
(J(\xp_i)\tensor 1)\cdot w_a =b' (\xp_i\tensor 1)\cdot w_a,\quad
 (1\tensor J(\xp_i))\cdot w_a = b (1\tensor\xp_i)\cdot w_a.
 \eqno(3.\num)
$$
While (3.5)
 may not hold in general, there are many situations
where
(3.5)  remains true even if the action of $Y(g)$ is non-trivial as
explained for (2.10).
 The examples (3.14)--(3.17) discussed later are indeed such cases
 except for the last two cases in (3.17).
\topinsert
{\ninerm \baselineskip=11pt
\noindent{ Table 1.\quad
 Explicit forms of  $g$-highest weights
in the decomposition (3.2). Here, $[34]\tensor [234]$ is a shorthand  for
$\xm_3 \xm_4 \cdot u_4 \tensor \xm_2 \xm_3 \xm_4 \cdot v_4$, etc.
}
 \medskip
\halign{# \hfil& \quad # \hfil 	\cr
\tabletoprule
module	&	highest weight vector	\cr
$V_{2\gL_4}$& $w_{2\cdot 4}=\bone \tensor
\bone \, ( \equiv u_4\tensor v_4)$	\cr
$V_{\gL_3}$& $w_3= [4] \tensor \bone- \bone\tensor [4]$\cr
$V_{\gL_1}$& $w_1=[43234]\tensor \bone - [3234]\tensor [4] +
	[234]\tensor [34] - [34]\tensor [234]$\cr
& $ \qquad\quad + [ 4]\tensor [ 3234] -\bone \tensor [43234] $\cr
$V_{\gL_4}$&$w_4= (2/3)[43231234]\tensor \bone
	-(1/3)[34231234]\tensor \bone -[3231234]\tensor [4]$ \cr
& $\qquad\quad + [231234]\tensor [34] -[31234]\tensor [234]
+[1234]\tensor [3234]$
\cr
& $\qquad\quad  + [3234]\tensor [1234] -[234]\tensor [31234]
+[34]\tensor [231234]$
\cr
&$\qquad\quad -[4]\tensor [3231234]
-(1/3)\bone\tensor[34231234]
+(2/3)\bone\tensor[43231234]$\cr
\tablebottomrule}
}
\endinsert
Since $w_a$ is $g$-highest, we have
$$
(\xp_i\tensor 1)\cdot w_a = - (1\tensor\xp_i)\cdot w_a
\ (\equiv y_{ia}\in V_{\imath(\gL_a)}) .
\eqno(3.\num)
$$
Using (2.4), (2.5), (3.5), and (3.6), we get
$$
J(\xp_i)\cdot w_a
	=  \left[ b'-b+\quar \bigl( c_2({\gL_a}) - c_2(\imath(\gL_a))\bigr) \right]
	y_{ia},
\eqno(3.\num)
$$
where $c_2(\gL)$ is the second Casimir of $V_{\gL}$.
{}From (3.7) we see that $w_a$ is $Y(g)$-highest if and only
 if\footnote{$\,^\dagger$}{A similar analysis shows that a
 $g$-lowest weight vector  of $V_{\gL_a}$ in (3.2) is $Y(g)$-lowest if and only
if
$ b' -b =  ( c_2(\imath^{-1}(\gL_a) )- c_2({\gL_a}) )/4$.}
$$
b' -b = \quar \bigl( c_2(\imath(\gL_a) )- c_2({\gL_a}) \bigr).
\eqno(3.\num)
$$
The relevant values of the second Casimir are
$
c_2({\gL_1})=18$,
$c_2({\gL_3})=24$, $c_2({\gL_4})=12$, and
$c_2({2\gL_4})=26$  (c.f.\ [\rMP]).

{\it Step 2).} Our next task is to determine the Drinfel'd polynomials of
the irreducible submodule generated by $w_a$
when $b'-b$ takes the value of (3.8). We give two independent ways to
do it.

The first way is to calculate
the eigenvalue of  $J(h_{a})$. Since the case $a=0$ is trivial, we concentrate
on the cases $a=1,3,4$.
Suppose we have the decomposition
$$
(h_a \tensor 1)\cdot w_a = y_{\parallel} + y_{\perp},
\quad y_{\parallel} \in V_{\gL_a}, \ y_{\perp}\in V_{\imath(\gL_a)}.
\eqno(3.\num)
$$
Let $\eta$ be the number defined by
$$
y_{\parallel} = \eta h_a \cdot w_a.
\eqno(3.\num)
$$
Then a similar calculation to (3.7) shows that
$$
J(h_a)\cdot w_a = \left( b+{\eta\over4}
\bigl( c_2(\imath(\gL_a) )- c_2({\gL_a})\bigr) \right)
h_a \cdot w_a.
\eqno(3.\num)
$$
(2.10) and (3.11) mean that the submodule generated by $w_a$
 is $W_a(b+\eta (c_2(\imath(\gL_a) )- c_2({\gL_a}))
/ 4 )$. In our example  a direct calculation  using  the explicit
 form of $w_a$
 in Table 1 shows that
$\eta = 1/2$ for  $a=1,3$ and $4$.

The second way is due to [\rBel] and more efficient in this example.
Let
$\gT = \sum_{p}I_p J(I_p)$.
Then
$$
\gD(\gT)=\gT\tensor 1 + 1 \tensor \gT
+\sum_p ( J(I_p) \tensor I_p + I_p\tensor J( I_p )).
\eqno(3.\num)
$$
Using (3.12), we have
$$
\gD(\gT) \cdot w_a =
\left( b +\half (b'-b) \right) c_2(\gL_a)  w_a .
\eqno(3.\num)
$$
This again means that
the submodule generated by $w_a$
 is $W_a(b+\half (c_2(\imath(\gL_a) )- c_2({\gL_a}))
/ 4 )$.
 This ends the description of Step 2.

Summarizing the results of Steps 1 and 2,
 we obtain the following homomorphisms of
 $Y(g)$-modules, or the
 ({\it injective}) {\it fusions}:
$$
\displaylines{
\hfill W_3(b )
 \hookrightarrow W_4(b+\quar)\tensor W_4(b-\quar), \quad
 W_1(b)
 \hookrightarrow W_4(b+1)\tensor W_4(b-1), \puteqno(3.\num)\cr
   W_4(b)
 \hookrightarrow W_4(b+{3\over2})\tensor W_4(b-{3\over2}) , \quad
\bC
 \hookrightarrow W_4(b+{9\over4})\tensor W_4(b-{9\over4}) .\cr }
 $$
 A hooked arrow $\hookrightarrow$ indicates that it is injective.
In the same way,
we get
$$
W_2(b)\hookrightarrow W_3(b+{1\over4})\tensor W_4(b-\half)
\hookrightarrow W_4(b+\half)\tensor W_4(b)\tensor W_4(b-\half).
\eqno(3.\num)
$$

For $Y(E_7)$  we get the following fusions, which produce
all the fundamental representations
from  the minimal representation $W_6(b)$:
$$
\eqalign{
&W_{6-a}(b )
 \hookrightarrow W_{7-a}(b+{1\over2})\tensor W_6(b-{a\over2}), \quad
\for\ a=1,2,3,
\cr
&W_1(b )
 \hookrightarrow W_6(b+{5\over2})\tensor W_6(b-{5\over2}), \quad
\bC
 \hookrightarrow W_6(b+{9\over2})\tensor W_6(b-{9\over2}) ,
 \cr
&
W_2(b)
 \hookrightarrow W_1(b+{\half})\tensor W_1(b-{\half}) ,
 \quad
 W_3(b)
 \hookrightarrow W_2(b+{1\over2})\tensor W_1(b-{1}) ,
\cr
&W_7(b)
 \hookrightarrow W_1(b+{3\over2})\tensor W_6(b-{2}) .\cr }
 \eqno(3.\num)
 $$
For $Y(E_8)$ all the fundamental representations
are produced from the minimal representation $W_1(b)$ by
the fusions,
$$
\eqalign{
&W_{a+1}(b )
 \hookrightarrow W_a(b+{1\over2})\tensor W_1(b-{a\over2}), \quad
\for\ a=1,2,3,4\cr
& W_7(b)
 \hookrightarrow W_1(b+3)\tensor W_1(b-{3}) , \quad
W_8(b)
 \hookrightarrow W_1(b+{5\over2})\tensor W_7(b-{3\over2}) ,\cr
&W_{6}(b)
 \hookrightarrow W_{7}(b+{1\over2})\tensor W_7(b-{\half}) , \quad
W_5(b)
 \hookrightarrow W_6(b+{\half})\tensor W_7(b-{1}) ,\cr
 &
 W_1(b)
 \hookrightarrow W_1(b+5)\tensor W_1(b-5), \quad
 \bC
 \hookrightarrow W_1(b+{15\over2})\tensor W_1(b-{15\over2}). \cr
 }
 \eqno(3.\num)
 $$
We also used the knowledge of the $R$-matrix from [\rCPiii] for the embedding
of $W_1(b)$ and $\bC$ in (3.17).

We call  a sequence of numbers
$[b_1,\cdots, b_n]$
a  {\it fusion content} of $W_a(b)$
if there is a fusion
$W_a(b)\hookrightarrow
W_{\rm min}(b+b_1)\tensor \cdots\tensor
W_{\rm min}(b+b_n)
$  [\rKNS]. $W_{\rm min}(b)$ is the minimal representation of $Y(g)$.
For instance
 $W_{\rm min}(b)=W_4(b)$ for $Y(F_4)$. Then
the results (3.14) and (3.15) can be written in
the form of  fusion contents of $W_1(b)$, $W_2(b)$, and $W_3(b)$
as
$[1,-1]$, $[\half, 0,-\half]$, and $[\quar, -\quar]$.
Either $[0]$ or $[{3\over2},-{3\over2}]$ is a fusion content
of $W_4(b)$. In general
for a given $W_a(b)$ its fusion
content is not necessary unique.
Fusion contents were used in [\rKNS]  for the study of the functional
relations among the transfer matrices of related
lattice models
without a proof.
Here we provided its  proof for $g=F_4$ case.
All the other fusion contents in [\rKNS] have been proved in the same way.
In Table 2 we summarize the minimal representations and
  fusion contents of  the fundamental representations for all $Y(g)$.
We use them in section 4.
\topinsert
{\ninerm \baselineskip=11pt
\noindent{ Table 2.\quad
Fusion contents of the fundamental representations of $Y(g)$.
The second column gives a number $a$ such that $W_a(b)$
is the minimal representation. In the third column
 $a:[b_1,\dots,b_n]$  represents that the $a$-th fundamental
representation has a fusion content $[b_1,\dots,b_n]$.
 $0$ is a trivial representation.
In $g=D_r$ and
$E_6$ two minimal representations are used.
 $[b_1,\overline{ b}_2]$ and  $[\overline{ b}_1,b_2]$  in $D_r$,
 for example, mean the fusions in
$W_r(b_1)\tensor W_{r-1}(b_2)$ and
$W_{r-1}(b_1)\tensor W_{r}(b_2)$, respectively.
See [\rKNS] for related information.
}
 \medskip
\halign{# \hfil& \hfil#\hfil& #\hfil\cr
\tabletoprule
$g$	& min&	\hfill fusion content \hfill	\cr
\tablerule
$A_r$	& $1$&
$a: [{a-1\over2},{a-3\over2},\dots,-{a-1\over2}]$,
 $(1\leq a\leq r)$,
$0:[{r\over2},{r-2\over2},\dots,-{r\over2}]$	\cr
$B_r$ & $r$
& $a: [{2r-2a-1\over4},-{2r-2a-1\over4}]$,
$(1\leq a\leq r-1)$, $r:[0]$,
 $0: [{2r-1\over4},-{2r-1\over4}]$\cr
$C_r$	&	$1$&
$a: [{a-1\over4},{a-3\over4},\dots,-{a-1\over4}]$,
 $(1\leq a\leq r)$,
$0:[{r+1\over4},-{r+1\over4}]$	\cr
$D_r$ & $r-1,r$
& $a:[{r-a-1\over2},-{r-a-1\over2}]$,
$[{\overline{r-a-1}\over2},-{\overline{r-a-1}\over2}]$,
$(1\leq a\leq r-2$,  $r-a$ even)\cr
& & $a:[{r-a-1\over2},-{\overline{r-a-1}\over2}]$,
$[{\overline{r-a-1}\over2},-{{r-a-1}\over2}]$,
$(1\leq a\leq r-2$, $r-a$ odd)\cr
&& $r-1: [\overline{ 0}]$, $r:[0]$,
 $0:[{r-1\over2},-{r-1\over2}]$,
 $[{\overline{r-1}\over2},-{\overline{r-1}\over2}]$,  ($r$ even)
 \cr
&&
 $0:[{r-1\over2},-{\overline{r-1}\over2}]$,
 $[{\overline{r-1}\over2},-{{r-1}\over2}]$,  ($r$ odd) \cr
 $E_6$ &$1,5$
 & $1: [0], [\overline{2},-\overline{2}]$,
  $2:[\half,-\half]$, $3:[1,0,-1]$, $[\overline{1},
   \overline{ 0}, -\overline{ 1}]$,
 $4:[\overline{\half},-\overline{\half}]$ \cr
 && $5:[\overline{ 0} ],[2,-2]$,
  $6:[{3\over2},-\overline{3\over2}], [\overline{3\over2},-{3\over2}]$,
 $0:[\overline{ 3},-3], [3,-\overline{ 3}]$
 \cr
 $E_7$ & $6$
 & $1:[{5\over2},-{5\over2}]$, $2:[3,2,-2,-3]$, $3:[{7\over2},
 {5\over2},{3\over2},-{3\over2},-{5\over2},-{7\over2}]$,
 $[{3\over2},\half,-\half,-{3\over2}]$ \cr
&& $4:[1,0,-1]$,
 $5:[\half,-\half]$, $6:[0]$, $7:[4,-1,-2],[2,1,-4]$,
 $0:[{9\over2},-{9\over2}]$
 \cr
 $E_8$ &$1$
 & $1:[0], [5,-5]$, $2:[\half,-\half]$,
 $3:[1,0,-1]$, $4:[{3\over2},\half, -\half,-{3\over2}]$\cr
& &$5:[2,1,0,-1,-2], [4,3,2,-2,-3,-4]$,
  $6:[{7\over2},{5\over2},-{5\over2},-{7\over2}]$, $7:[3,-3]$\cr
&& $8:[{9\over2},-{3\over2},-{5\over2}], [{5\over2},{3\over2},-{9\over2}]$,
 $0:[{15\over2},-{15\over2}]$
 \cr
 $F_4$ & $4$
 & $1:[1,-1]$, $2:[\half,0,-\half]$, $3:[\quar,-\quar]$,
 $4:[0],[{3\over2},-{3\over2}]$, $0:[{9\over4},-{9\over4}]$
 \cr
 $G_2$ & $2$
 & $1:[{1\over6},-{1\over6}]$, $2:[0], [{2\over3},-{2\over3}]$,
 $0:[1,-1]$ \cr
\tablebottomrule}
}
\endinsert

\subsec{Multiplicity non-free case}
The content of this subsection will not be used later, so that
the reader
may skip it.

Even  when $g$-highest weight vectors in a tensor product have
multiplicity greater than one, one can still apply  the same technique
with a slight modification.
Again the problem is to find a $Y(g)$-highest weight vector,
which may occur at a certain value of $b'-b$,
among the $g$-highest weight vectors of a given weight.
But this time  we  need to
 choose a particular linear combination of them so that
it becomes $Y(g)$-highest. Below we show in an example that this
 procedure is indeed possible if we
know the action of $J(x)$ on these $g$-highest weight vectors.

Let $g=G_2$. We consider
 the tensor product
$W_2(b')\tensor W_1(b)$. As $g$-modules [\rDri]
$$
W_1(b)\simeq V_{\gL_1} \oplus \bC,
\quad
W_2(b')\simeq V_{\gL_2}.
\eqno(3.\num)
$$
Thus as a $g$-module,
$$
W_2(b')\tensor W_1(b)
\simeq
V_{\gL_1+\gL_2}\oplus V_{2\gL_2} \oplus 2 V_{\gL_2}.
\eqno(3.\num)
$$
Therefore $g$-highest weight vectors with weight $\gL_2$ is not multiplicity
free.
We write an $g$-highest weight vectors of $V_{\gL_1}$ and $V_{\gL_2}$
 in the left hand side of (3.19) as $v_1$ and $u_2$.
Since $V_{\gL_1}$ is the adjoint representation, we  identify
$V_{\gL_1}$ with $g$. Explicitly the correspondence is
$$
\eqalign{
&[2221]	\leftrightarrow 6 \xp_1,
\quad
[1221]	\leftrightarrow -2 \xp_2,
\cr
&[12221]	\leftrightarrow -6 h_1,
\quad
[21221]	\leftrightarrow 2 h_2,
\cr}
\eqno(3.\num)
$$
and so on   (c.f.\ [\rFH]),
where we used the same notation  as in Table 1, namely,
$[2221]=\xm_2 \xm_2 \xm_2 \xm_1 \cdot v_1$, etc.
Under our normalization $(\ga_1,\ga_1)=2$
the action of $J(x)$ on $W_1(b)$ is defined by [\rDri,\rCPiii]
$$
J(x)\cdot (y,\gl)=(-{10\over9}\gl x, (x,y)) + bx\cdot (y,\gl),
\quad y\in g, \gl\in \bC .
\eqno(3.\num)
$$
The action of $J(x)$ on $W_2(b')$ is given by that of $b'x$.

Let us choose independent $g$-highest weight vectors,
 $w_2$ and $w'_2$,
of
weight $\gL_2$
in $W_2(b')\tensor W_1(b)$ as
$$
\eqalign{
w_2&=3[12212]\tensor (\bone,0)
-3 [2212]\tensor ([1],0)
+2  [212]\tensor ([21],0)
-[12]\tensor ([221],0)	\cr
&\qquad +[2]\tensor ([1221],0)
-2\cdot  \bone \tensor ([21221],0)
+ \bone\tensor ([12221],0) ,\cr
w'_2&= \bone \tensor (0, 1) .\cr
}
\eqno(3.\num)
$$
Then $(\xp_i\tensor 1)\cdot w_2
\in V_{\gL_1+\gL_2} \oplus V_{2\gL_2}$ and $(\xp_i\tensor 1)\cdot w'_2=0$.
Thus this time the tensor product graph
$$
\matrix{
 &\gL_1 + \gL_2 & \cr
\hfill\nearrow&& \nwarrow\hfill \cr
2\gL_2\ \ & \longleftarrow &\gL_2\oplus   \gL_2 \cr
}
\eqno(3.\num)
$$
does not immediately tell us whether there exits a $Y(g)$-highest weight
vector with weight $\gL_2$ or not. A direct calculation, using (3.20--22),
 however
shows that for $i=1$ and $2$
$$
\eqalign{
J(\xp_i)\cdot (w_2 + \ga w'_2) &=
[ b'-b-\quar ( c_2(\gL_1 + \gL_2) - c_2(\gL_2) ) -\ga ] y_i \cr
&\qquad +
[ b'-b-\quar ( c_2(2 \gL_2) - c_2(\gL_2) ) +{3\over4}\ga] y'_i \cr
}
\eqno(3.\num)
$$
for some nonzero vectors $y_i\in V_{\gL_1+\gL_2}$
 and $y'_i \in  V_{2\gL_2}$.
Therefore the vector $w_2 + \ga w'_2$ is $Y(g)$-highest if and only if
$$
\eqalign{
\ga&={1\over7} (c_2(2\gL_2) - c_2( \gL_1 + \gL_2)),
\cr
b'-b &= {1\over4}\left( {3\over7}c_2(\gL_1 + \gL_2) +
{4\over7} c_2(2\gL_2) - c_2(\gL_2) \right) .	\cr
}
\eqno(3.\num)
$$
Notice that the value $b'-b$ in (3.25) is not
the difference of two second Casimirs over 4 any more.
A similar but a little more nontrivial
 calculation like (3.9--11) shows
$$
J(h_2)\cdot (w_2 + \ga w'_2)
=(b -{\ga\over4}) h_2\cdot w_2+ (b'+{10\over9\ga}) h_2 \cdot \ga w'_2
\eqno(3.\num)
$$
for $\ga$ and  $b'-b$ satisfying  (3.25).  Substituting the values
$c_2(\gL_1+\gL_2)=14$,  $c_2(2\gL_2)=28/3$, $c_2(\gL_1)=8$,
and
$c_2(\gL_2)=4$, we have $\ga=-{2\over3}$ and $b'-b={11\over6}$.
Thus we obtain the following fusion
$$
W_2(b) \hookrightarrow W_2(b+{5\over3})
\tensor W_1(b-{1\over6}) .
\eqno(3.\num)
$$
\subsec{Duality}

The existence of a dual (contragradient)
representation is guaranteed by
the antipode of $Y(g)$.
We use a theorem
 from [\rCPiii], Proposition 2.17: If $V_{\gL_{\overline{ a}}}$ is the dual
 representation of $V_{\gL_a}$ for $g$, then
$W_{\overline{ a}}(b-{\dcn\over 2})$ is dual to $W_{a}(b)$ in the following
sense:
$$
\eqalignno{
{\rm i).} &\quad\bC\hookrightarrow W_{ a}(b)\tensor W_{\overline{ a}}
(b-{\dcn\over 2}). &(3.\num{\rm a})\cr
{\rm ii).} &
\quad \hbox{For any finite-dimensional representations $U$ and $V$
of $Y(g)$} \cr
&
\qquad \Hom_{Y(g)}(U, W_{a}(b)\tensor V)
\simeq \Hom_{Y(g)}(W_{\overline{ a}}(b-{\dcn\over 2})\tensor U, V) .\quad
& (3.\numadd{\rm b})\cr
}
$$
(3.28b), in particular, means if there is a nontrivial homomorphism
 $U\rightarrow W_{a}(b)\tensor V$,
there is also a nontrivial  homomorphism
 $W_{\overline{ a}}(b-{\dcn\over 2})\tensor U \rightarrow V$.

So, if $g\neq A_r, D_r, E_6$, then $W_a(b-{\dcn\over 2})$ is dual
to $W_a(b)$. For
 $g= A_r, D_r, E_6$ the  diagram automorphism of $g$
 is also an automorphism of $Y(g)$. Thus, a fusion
$$
W_{{a}_3}(b_3)\hookrightarrow
W_{a_1}(b_1)\tensor  W_{a_2}(b_2)
\eqno(3.\num)
$$
is equivalent to its conjugate fusion
$$
W_{\overline{a}_3}(b_3)\hookrightarrow
W_{\overline{a}_1}(b_1)\tensor  W_{\overline{a}_2}(b_2).
\eqno(3.\num)
$$
On the other hand, by tensoring $W_{\overline{a}_3}(b_3-
{\dcn\over2})$ to (3.29) from the right, then
using (3.28a), we have
$$
\bC\hookrightarrow
W_{a_1}(b_1)\tensor  W_{a_2}(b_2)\tensor
W_{{a}_3}(\overline{b}_3-{\dcn\over2}).
\eqno(3.\num)
$$
{}From (3.31) and the remark after (3.28b), we have a nontrivial
homomorphism,
$$
W_{\overline{a}_2}(b_2-{\dcn\over2})\tensor
W_{\overline{a}_1}(b_1-{\dcn\over2})
\rightarrow W_{\overline{a}_3}(b_3-{\dcn\over2}).
\eqno(3.\num)
$$
Since $W_{\overline{a}_3}(b_3)$ is $Y(g)$-irreducible,
this homomorphism is surjective by Schur's lemma.
Thus (3.29) is also equivalent to the following {\it surjective fusion}:
$$
W_{{a}_2}(b_2)\tensor
W_{{a}_1}(b_1)
\rightarrow W_{{a}_3}(b_3) .
\eqno(3.\num)
$$

\newsec{Applications to an $S$-matrix model with the $Y(g)$ symmetry}
\eqnumber=0
\subsec{Extension by the Poincar\'e algebra and mass formula}
We recall [\rBer] that $Y(g)$ defined by (2.1) admits a non-trivial extension
by the on-shell two-dimensional Poincar\'e algebra ${\cal P}$,
$$
[L,P^0]=P^1,\quad [L,P^1]=P^0,\quad [P^0,P^1]=0,
\quad (P^0)^2-(P^1)^2=m^2,
\eqno(4.\num)
$$
where $L, P^0$,  $P^1$ are the Lorentz boost, the energy, the momentum
operators, and $m$ is a central element.
 The only non-trivial relation between $Y(g)$ and
${\cal P}$ is
$$
[L,J(x)]=\gc^{-1} x,
\eqno(4.\num)
$$
where $\gc$ is a coupling constant between the two algebras.
Then one can easily check (see   Appendix)
that (4.2) is compatible with the relations
(2.1) and (4.1).
The comultiplication for $X \in {\cal P}$
 is defined as
$$
\gD(X)=X\tensor 1 + 1 \tensor X.
\eqno(4.\num)
$$
Let us write this extended Hopf algebra as $PY(g)$.

Remarkably,  like the Virasoro algebra in an affine Lie algebra,
the operator $L$ is already built into $Y(g)$. To see this, we use
 a one-parameter family of automorphisms $T_b$ ($b \in \bC$)
 of $Y(g)$ such that
[\rDri]
$$
T_b : x \mapsto x,\quad T_b: J(x) \mapsto J(x) + b x.
\eqno(4.\num)
$$
The pull-back  of $W_a(0)$ by
the map $T_b : Y(g)\rightarrow Y(g)$
 induces a
$Y(g)$-module, which is isomorphic to
$W_a(b)$ due to Prop.\ 2.14 in [\rCPiii].
Thus
 one can define a representation of $PY(g)$ on  $W_a(b)$ as
$$
\eqalign{
&J(x)=J(x)_{{\rm for\ }b=0}+b x, \quad
L={1\over\gc} {\partial\over\partial b}, \cr
& P^0=m_a\cosh(\gc b+d_a),
\quad P^1 =m_a \sinh (\gc b+d_a). \cr
}
\eqno(4.\num)
$$
The constants $m_a$ and  $d_a$ are arbitrary  at this moment.
This representation
is a generalization of the ideas of [\rBer,\rBel,\rMac].

Now, following  [\rBel],
suppose a quantum field theory possesses the $Y(g)$ symmetry.
Furthermore suppose that there are $r\, (=\rank\,  g)$ particles $a=1,\dots,
r$,
 and  each particle
$a$
consists of a multiplet belonging to the
fundamental representation $W_a(b)$
with the mass $m_a>0$.
The $G\tensor G$-invariant nonlinear $\gs$-model
 is   an example   [\rOW,\rORW].
Let us assume that a bound state occurs when there is a
corresponding, either injective or surjective
 fusion (we return to this point in
 section 4.2 again).
 Then
let us  see what the results (3.14) and (3.15)
 in the previous section imply
for the $Y(F_4)$ case.
The energy and momentum conservations (4.3)
give constraints
on $m_a$ and $d_a$, which are summarized into a set of equations,
$$
\eqalign{
m_4 (e^{\gc} + e^{-\gc}) &= m_1e^{\pm(d_1-d_4)},\cr
m_4 (e^{\gc/2} +1+ e^{-\gc/2}) &= m_2e^{\pm(d_2-d_4)},\cr
m_4 (e^{\gc/4} + e^{-\gc/4}) &= m_3 e^{\pm(d_3-d_4)},\cr
m_4 (e^{3\gc/2} + e^{-3\gc/2}) &= m_4,\cr
m_4 (e^{9\gc/4} + e^{-9\gc/4}) &= 0.\cr
}
\eqno(4.\num)
$$
In the last equation of (4.6) we
 required the mass $m_0$ of the trivial representation
$\bC$ to be zero.
It is immediate to see that it has a solution only if $d_a$ is independent
of $a$. From now on
 we set $d_a=0$  without losing generality. Thus the {\it rapidity}
 $\gth_a$
of $W_a( b)$ is identified with $\gc b$.
Requiring $m_a > 0$,   (4.6)  determines  $\gc$ and $m_a$ as
$$
\displaylines{
\hfill\gc=\pm {2\pi i \over 9}\quad \mod\ 8\pi i ,\puteqno(4.\num)\cr
m_1 = (2\cos {2\pi\over9}) m_4,\quad
m_2 =(1+2\cos {\pi\over9} ) m_4,\quad
m_3 = ( 2\cos {\pi\over18}) m_4. 	\cr
}
$$
This agrees with the mass formula
in [\rOW], upon correcting the obvious typographical errors therein.

\par
It is straightforward to repeat this calculation for all $Y(g)$.
  Except for $Y(G_2)$ the requirements $m_a>0$ and $m_0=0$ discretize
  the coupling constant $\gc$ as
$$
\gc = \pm {2 \pi i\over \dcn}\quad \mod\ 4d \pi i,
\quad
d=\cases{
3 &for $G_2$,\cr
2&for $B_r, C_r$, and  $F_4$,\cr
1&otherwize.\cr
}
\eqno(4.\num)
$$

\topinsert
{\ninerm
\noindent{ Table 3.\quad
 The  dual Coxeter number of   $g$.
}
 \medskip
\halign{# \hfil&\hfil  #\hfil &\hfil #\hfil&\hfil #\hfil&\hfil #\hfil&
\hfil #\hfil& \hfil#\hfil&\hfil #\hfil& \hfil#\hfil&\hfil #\hfil\cr
\tabletoprule
$g$	&	$A_r$ &$B_r$ &$C_r$ &$D_r$ &$E_6$ &
$E_7$ &$E_8$ & $F_4$ &$G_2$ 	\cr
dual Coxeter number	&	$r+1$ &$2r-1$ &$r+1$ &$2r-2$ &$12$ &
$18$ &$30$ & $9$ &$4$ 	\cr
\tablebottomrule}
}
\endinsert
\noindent
The value of the dual Coxeter number  $\dcn$  is listed in Table 3.
The ratios of  the masses $m_a$ are unique under (4.8).
Let    $[b_1,\cdots, b_n]$ be a fusion content of $W_a(b)$ in Table 2,
and  $m_{\rm min}$ be  the mass of  the minimal
representation $W_{\rm min}(b)$.
Then
the mass $m_a$ is given in a unified way:

\proclaim Mass formula.
$$
m_a =
 m_{\rm min} \cdot \sum_{j=1}^n e^{2 b_j \pi i /\dcn} .
\eqno(4.\num)
$$
\par
\noindent
In the case $Y(G_2)$, besides the above, there is one more
positive mass solution,
$$
\gc = \pm {5\pi i\over2}\quad \mod\ 12\pi i,\quad
m_1 = (2\cos {5\pi\over12}) m_2.
\eqno(4.\num)
$$
It is nothing
 surprising
that (4.9) with  the data of the fusion contents in Table 2
reproduces the
mass formula  in [\rOW],\footnote{$^\dagger$}{
In [\rOW] the mass $m_3$ (in their notation) of  $E_6$
should read as $(\sqrt{3}+1)m_1$, and $m_5$ and $m_6$ of $E_8$
 should be interchanged.}
 because  the  equations in  (4.6) are equivalent to the
ones
used in  the
 bootstrap method  [\rZZ,\rOW,\rORW].
 We, however,  clarified  here how the two-dimensional
 kinematics is compatible with $Y(g)$.

\subsec{Triangle relation and physical strip condition}

So far we have considered only the positivity condition of mass for
the restriction of the coupling constant $\gc$.
In [\rBer] it was mentioned that only the value $\gc=2\pi i/ \dcn$
 ensures the crossing symmetry of
 the $S$-matrix.
 (If there is a crossing symmetric
$S$-matrix at $\gc=2\pi i/ \dcn$, however,
then there  is  also one  at $\gc=-2\pi i/ \dcn$.)
Below we shall find the values $\gc=\pm 2\pi i/ \dcn$
using a different argument.

Let $\overline{a}$ denote the antiparticle of $a$.
General theory of $S$-matrices tells  that
if there is a bound state $\overline{ a}_3$ in the process
$a_1+a_2 \rightarrow \overline{ a}_3$, then the $S$-matrix,
as a function of the difference of the rapidities, say,
$\gth_1-\gth_2$, has a pole at a point
 $i u_{a_1 a_2}^{\overline{ a}_3}$ with
 $$
 0< u_{a_1 a_2}^{\overline{ a}_3} < \pi .
 \eqno(4.\num)
 $$
We  call   $u_{a_1 a_2}^{\overline{ a}_3}$
the fusion  angle, and (4.11)
the { physical strip condition}.
On the other hand the energy and momentum conservations, like (4.6),
give a relation,
$$m_{a_3}^2=m_{a_1}^2+m_{a_2}^2+2m_{a_1}m_{a_2}\cos
u_{a_1 a_2}^{\overline{ a}_3} ,
\eqno(4.\num)
$$
among the masses and the fusion angle.
Let us further {\it assume} that there are also the binding
processes
$a_2+a_3 \rightarrow \overline{ a}_1$ and
$a_3+a_1 \rightarrow \overline{ a}_2$.
Then using (4.11) and (4.12), we have
 the  {\it triangle relation}:
$$
u_{a_1 a_2}^{\overline{a}_3} +
u_{a_2 a_3}^{\overline{a}_1}+
u_{a_3 a_1}^{\overline{a}_2} =2\pi .
\eqno(4.\num)
$$
See  [\rZZ, \rZam] for details.
We shall show that  (4.13) follows from the representation theory
of $Y(g)$ if and only if we choose $\gc=\pm 2\pi i/\dcn$ among
the  possible values in (4.8) and (4.10).

Below we must use one unproven fact:

\proclaim Assumption.
In our convention of the comultiplication (2.4),
 there is an injective fusion,
$$
W_{a_3}(b_3) \hookrightarrow
W_{a_1}(b_1)\tensor W_{a_2}(b_2),
\eqno(4.\num)
$$
only if\/ $b_1 > b_2$.
\par
 This is supposed to be true in the following reason.
  In general, if $b_1 < b_2 $, then any non-trivial submodule
  in $W=W_{a_1}(b_1)\tensor W_{a_2}(b_2)$
  has the highest weight $\gL_{a_1} + \gL_{a_2}$.
  In other words $W_{a_3}(b_3)$ may be isomorphic to
  a quotient, but not to a submodule of $W$.
Yet we do not know a general proof of the
statement at this moment.

Now suppose there is a fusion
$$
W_{\overline{a}_3}(b_3)\hookrightarrow
W_{a_1}(b_1)\tensor  W_{a_2}(b_2).
\eqno(4.\num)
$$
By an  argument similar to the one from (3.29) to (3.32),
we can show the existence of  another fusion
$$
W_{\overline{a}_1}(b_1-{\dcn\over2})\hookrightarrow
W_{a_2}(b_2)\tensor  W_{a_3}(b_3-{\dcn\over2}).
\eqno(4.\num)
$$
Repeating the  procedure once more, we also have
$$
W_{\overline{a}_2}(b_2-{\dcn\over2})\hookrightarrow
W_{a_3}(b_3-{\dcn\over2})\tensor  W_{a_1}(b_1-\dcn).
\eqno(4.\num)
$$
The differences of the rapidities in (4.15)--(4.17) are
$$
\gth_{a_1 a_2}^{\overline{a}_3}=\gc(b_1-b_2),
\quad
\gth_{a_2 a_3}^{\overline{a}_1}=\gc(b_2-b_3+{\dcn\over2}),
\quad
\gth_{a_3 a_1}^{\overline{a}_2}=\gc(b_3-b_1 +{\dcn\over2}).
\eqno(4.\num)
$$
Therefore
$$
\gth_{a_1 a_2}^{\overline{a}_3} +
\gth_{a_2 a_3}^{\overline{a}_1}+
\gth_{a_3 a_1}^{\overline{a}_2} =\gc \dcn .
\eqno(4.\num)
$$
So far we have not yet used our assumption on (4.14).
As an example, let $g=G_2$. From  (3.27) we have
$\gth_{12}^{\overline{2}}=\gth_{21}^{\overline{2}}=(11/6)\gc$.
Also the fusion content of
$W_1(b)$ in Table 2 gives $\gth_{22}^{\overline{1}}=(1/3)\gc$.
Thus $\gth_{12}^{\overline{2}}+\gth_{22}^{\overline{1}}+
\gth_{21}^{\overline{2}}=4\gc=\dcn\gc$.

Now let $\gc={2\pi i/ \dcn}$, and let
 $u_{a_1 a_2}^{\overline{a}_3}=-i
\gth_{a_1 a_2}^{\overline{a}_3}$, etc. Then from the assumption on
(4.14) we have $u_{a_1 a_2}^{\overline{a}_3},
u_{a_2 a_3}^{\overline{a}_1},
u_{a_3 a_1}^{\overline{a}_2} > 0$, and from (4.19) they satisfies
the triangle relation (4.13). Furthermore
it is an exercise of elementary geometry  to show
$$
u_{a_1 a_2}^{\overline{a}_3} ,
u_{a_2 a_3}^{\overline{a}_1},
u_{a_3 a_1}^{\overline{a}_2} < \pi
\eqno(4.\num)
$$
{}from (4.12) and (4.13). We can repeat the same argument for
$\gc=-{2\pi i/ \dcn}$ using the duals of (4.15)--(4.17).
We conclude that if $\gc={2\pi i/ \dcn}$
(resp.\ $\gc=-{2\pi i/ \dcn}$), then any
injective (resp.\ surjective) fusion  occurs in
the region $0< \Im\, \gth_1 - \gth_2 < \pi $.

Thus $Y(g)$ representation theory
guarantees that i)
all the fusion angles among fundamental representations
are automatically inside the physical strip,
ii) if there is a binding process $a_1+a_2 \rightarrow \overline{ a}_3$,
then there are also the others, $a_2+a_3 \rightarrow \overline{ a}_1$ and
$a_3+a_1 \rightarrow \overline{ a}_2$.

\subsec{Integrals of motion}

It is commonly believed that in classical and quantum
 field theories the integrability
and the existence of the infinite numbers of
integrals of motion (IM) commuting with
each other are synonymous. It is an interesting
problem to realize the corresponding conserved
currents as  composite operators of the non-local
conserved current in [\rBer].

Suppose we have an IM $I_s$ of spin $s$, i.e., $[L,I_s]=sI_s$.
Then the charge of $I_s$ on
the particle $a$ with rapidity $\gth_a$
is $\go_a e^{s\gth_a}$ for some number $\go_a$ We
call $\go_a$ the scalar charge.
In [\rZam] it was pointed out that the conservation of
the $I_s$ charge gives a strong constraint among the
the spin $s$ and the fusion angles.
When $g$ is simply-laced,
the mass spectrum of (4.9), thus the fusion angles, are identical to the one
in the affine Toda field theory
where the situation is well-known [\rFZ--\rFF];
an IM with spin $s$
exists if and only if $s$ is  an exponent of $g$ modulo the
Coxeter number.

We set $\gc=\pm 2\pi i /\dcn$ in the rest of the paper.
Let us first consider $g=F_4$, for example.
The conservation of the $I_s$ charge under the fusion $W_4(b)
\hookrightarrow W_4(b+{3\over2})\tensor
W_4(b-{3\over2})$ in (3.14) gives an consistency equation for $s$,
$$
e^{3s\gc/2} +e^{-3s\gc/2} = 2\cos {s\pi\over 3}=1.
\eqno(4.\num)
$$
This excludes the spins $s=0,2,3,4$ mod 6. Remembering that
the exponents of $F_4$ are $1,5,7,11$ and the Coxeter number
is $12$, we  have already obtained a selection rule:
\proclaim Selection rule.
If an $S$-matrix has the $Y(g)$ symmetry, then
 an integral of motion with spin $s$
 exists  only if $s$ is  an exponent of $g$ modulo the
Coxeter number.

Of course this argument does not necessarily guarantee the existence
of an IM at the allowed spins.
The systematic absence of IMs at these particular values,
 however,  suggests that  IMs really
exist at the allowed spins.
Below we give the derivation of the above selection rule
for all the other algebras. The  Coxeter number
and the exponents of $g$
are listed in Table 4.
\topinsert
{\ninerm
\noindent{ Table 4.\quad
The Coxeter number and the exponents of $g$.}
 \medskip
\halign{\hfil# \hfil& \hfil #\hfil & #\hfil\cr
\tabletoprule
$g$	&	Coxeter number & \hfil exponents 	\cr
\tablerule
$A_r$	& $r+1$& $1,2,\dots,r$ \cr
$B_r$	& $2r$& $1,3,5,\dots,2r-1$ \cr
$C_r$	& $2r$& $1,3,5,\dots,2r-1$ \cr
$D_r$	& $2r-2$& $1,3,5,\dots,2r-3,r-1$ \cr
$E_6$	& $12$& $1,4,5,7,8,11$ \cr
$E_7$	& $18$& $1,5,7,9,11,13,17$ \cr
$E_8$	& $30$& $1,7,11,13,17,19,23,29$ \cr
$F_4$	& $12$& $1,5,7,11$ \cr
$G_2$	& $6$& $1,5$ \cr
\tablebottomrule}
}
\endinsert

$A_r$: A fusion content of  $\bC$ is
$[{r\over2},{r-2\over2},\dots,-{r\over2}]$. This leads to a constraint
$$
\sum_{j=0}^r e^{2 sj \pi i / (r+1)} =0.
\eqno(4.\num)
$$
Thus $s=0$ mod $r+1$ are excluded.

$B_r$: A fusion content of  $\bC$ is
$[{2r-1\over4},-{2r-1\over4}]$. This leads to a constraint
$$
\cos {s\pi \over2}=0.
\eqno(4.\num)
$$
Thus $s={\rm even}$ are excluded.

$C_r$: A fusion content of  $\bC$ is
$[{r+1\over4},-{r+1\over4}]$. This leads to the same constraint
as (4.23).
Thus  $s={\rm even}$ are excluded.

$D_r$ ($r$ even): A fusion content of  $\bC$ is
$[{r-1\over2},-{r-1\over2}]$. This leads to the same constraint
as (4.23).
Thus  $s={\rm even}$ are excluded.

$D_r$ ($r$ odd): This and $E_6$ cases require two-steps examination.
There are two minimal representations $W_r(b)$ and $W_{r-1}(b)$,
the spin and conjugate spin representations.
 Let $\go $ and $ \go' $ be
the scalar charges of an IM $I_s$ on the particles
$r$ and $r-1$.
For $r-a$ even, there are two fusion contents  of $W_a(b)$,
$[{r-a-1\over2},-{r-a-1\over2}]$ and
$[{\overline{r-a-1}\over2},-{\overline{r-a-1}\over2}]$.
They lead to a consistency equation
$$
(\go-\go') \cos {s(r-a-1)\pi \over 2r-2}=0.
\eqno(4.\num)
$$
If $s=r-1$ mod $2r-2$, it gives no constraint for $\go$ and $\go'$.
If $s\neq r-1$ mod $2r-2$,  we have $\go=\go'$.
The fusion content of $\bC$, $[{r-1\over 2}, - {\overline{r-1}\over2}]$,
leads to another constraint
$$
\go e^{s\pi i/2} + \go'  e^{-s\pi i/2} =0.
\eqno(4.\num)
$$
If $s=r-1$ mod $2r-2$, then $\go=-\go'$ satisfies (4.25).
Therefore there is no constraint for $s$.
If $s\neq r-1$ mod $2r-2$, where  $\go=\go'$,
(4.25) excludes
 $s={\rm even}\neq r-1$ mod $2r-2$.

$E_6$:
There are two minimal representations $W_1(b)$ and $W_{5}(b)$.
 Let $\go $ and $ \go' $ be
the scalar charges of an IM $I_s$ on the particles
$1$ and $5$.
The fusion contents, $[2,-2]$ of $W_5(b)$ and
$[\overline{ 2}, -\overline{ 2}]$ of $W_1(b)$, give  equations
$$
 \go' = \left( 2\cos {s\pi \over 3} \right) \go,
\quad
 \go =  \left(  2\cos {s\pi \over 3} \right)\go',
 \eqno(4.\num)
$$
which exclude $s=0$ mod 3, and set
$\go=\go'$ for $s=1,5$ mod 6
and $\go=-\go'$ for $s=2,4$ mod 6.
Next compare  two fusion contents of $W_6(b)$,
$[{3\over2},-\overline{3\over2}]$ and $[\overline{3\over2},-{3\over2}]$.
This does not give any constraint for $s=1,5$ mod 6, but
gives a constraint $\sin {s\pi\over4}=0$ for $s=2,4$ mod 6.
This excludes $ s=2,10$ mod 12. Hence we obtain the rule.
Alternatively we can get the same result as follows:
The fusion contents $[2,-2]$ of $W_5(b)$ and
$[\overline{1},\overline{0},-\overline{1}]$
of $W_3(b)$ mean another fusion content
$[3,-1,2,-2,1,-3]$ of $W_3(b)$.
Comparing this with $[1,0,-1]$, we have
a consistency equation
$$
\cos{ s \pi \over 2} + \cos{s \pi \over 3} = {1\over 2}.
\eqno(4.\num)
$$
$s$ is a solution of (4.27) if and only if it is
an exponent of $E_6$ mod 12.

$E_7$: Comparing two fusion contents of $W_3(b)$,
$[{7\over2},{5\over2},{3\over2},-{3\over2},-{5\over2},-{7\over2}]$
and $[{3\over2},{1\over2},-{1\over2},-{3\over2}]$,
 we have
an equation
$$
\cos {7s\pi\over18}+
\cos {5s\pi\over18}-
\cos {s\pi\over18}=0.
\eqno(4.\num)
$$
$s$ is a solution of (4.28) if and only if  it is an exponent of $E_7$ mod
$18$.
Thus the other spins are excluded.

$E_8$: Comparing two fusion contents of $W_5(b)$,
$[4,3,2,-2,-3,-4]$
and $[2,1,0,-1,-2]$,
 we have
an equation
$$
\cos {4s\pi\over15}+
\cos {3s\pi\over15}-
\cos {s\pi\over15}=\half.
\eqno(4.\num)
$$
$s$ is a solution of (4.29)  if and only if  it is an exponent of $E_8$ mod
$30$.
Thus the other spins are excluded.

$G_2$: A fusion content of  $W_2(b)$ is
$[{2\over3},-{2\over3}]$. This leads to the same constraint
as (4.21).
Thus $s=0,2,3,4$ mod 6 are excluded.

To sum up, the selection rule follows only from the consistency
of the fusion data in a remarkably simple way.

Like (4.9), when there is an IM $I_s$ of spin $s$,
the ratios of the scalar charges of $I_s$ are
 given in terms of a fusion content of $W_a(b)$
as
$$
\go_a = \go_{\rm min} \cdot \sum_{j=1}^n e^{ 2 s b_j \pi i /\dcn}
\quad {\rm for\ } g\neq D_r, E_6 .
\eqno(4.\num)
$$
Here $\go_{\rm min}$ is the scalar charge on the particle
corresponding to the minimal representation.
 For $g= D_r$ and  $E_6$, by (4.25) and (4.26), (4.30) is modified as
 $$
\go_a = \go_{\rm min} \cdot \left( \sum_{j=1}^n
e^{ 2 s b_j \pi i /\dcn}
- (-1)^{s} \sum_{j=1}^{n'} e^{ 2 s b'_j \pi i /\dcn} \right),
\eqno(4.\num)
$$
where    $\go_{\rm min}$ is
the scalar charge of the particle $r$ (resp. the particle 1) for $D_r$ (resp.
$E_6$),
and
$[b_1, \dots, b_n, \overline{b'}_1,\dots, \overline{b'}_{n'}]$
is a  fusion content of $W_a(b)$.
It is easy to check the property $\go_a = -(-1)^s
 \go_{\overline{a}}$ from (4.31) [\rFZ]. For a simply-laced $g$
 the numbers
 $\go_a$ ($a=1,\dots, r$) constitute an eigenvector
 of the Cartan algebra of $g$ with eigenvalue $2-2\cos(s\pi /\dcn)$
 as expected from  the result in the corresponding affine Toda theory
for $g^{(1)}$ [\rBCDS].
For a non-simply-laced $g$
 those numbers $\go_a$
 coincide with the scalar charges of IMs in
the classical affine Toda theory for the {\it dual} of $g^{(1)}$,
which is a twisted affine Lie algebra
  [\rBCDS,\rDGZa,\rDori]\footnote{$\,^\dagger$}{
We thank P.\ Dorey for making us aware of this point.}.

\par

\newsec{Conclusion}

We  showed how  representation theory of $Y(g)$ in
 [\rDri,\rDrii-\rCPiii] works  well
in    the study of  the fusion procedures,
the mass spectrum, the triangle relation, and integrals of motion
when the $S$-matrix of a massive model has the $Y(g)$ symmetry.
We observed that the Yangian
perfectly integrates the $S$-matrix theory into
itself.
This is why the traditional bootstrap
approach in [\rZZ,\rORW] worked  so consistently.
Also this reminds us what the affine Lie algebra did for the
Wess-Zumino-Witten conformal field theory.
This point of view has been already stated in [\rBer].
Technically  the key to go beyond the traditional
bootstrap and the $R$-matrix method was
the  identification of a  one-particle state with
a (fundamental) highest weight representation parametrized
by the Drinfel'd polynomials.

Finally it is important to understand  any intrinsic relation
between the fusion of the Yangian representations and the one
described by the Coxeter element [\rDorii].

\bigbreak\bigskip\bigskip\centerline{{\bf Acknowledgements}}\nobreak
We would like to thank  A.\ Kuniba for pointing out to me the importance of the
Drinfel'd polynomials, and thank to him and J.\ Suzuki for the
collaboration which
leads to the work presented here.
We thank  J.\ Beck, I.\ Cherednik, S.\ Cordes, P.\ Dorey,
M.T.\ Grisaru, N.\ MacKay, P.\ Mathieu,
  P.B.\ Wiegmann
for interesting discussion and correspondence,
M.\ Bershadsky and C.\ Vafa for their warm hospitality. This work is
supported by JSPS fellowship, NSF grant PHY 92-18167, and Packard
fellowship.

\vskip1cm
\noindent
{\bf Appendix. Compatibility between (2.1) and (4.2)}
\eqnumber=0
\def\rA{{\rm A}}
\par
The relation (4.2) must be compatible with the relations (2.1) and (4.1).
The compatibility with (4.1) is clear.

To check the compatibility with (2.1),
 we have only to apply $\gc[ L,\cdot]$ to
the relations in (2.1) and see the equality.
 The case (2.1a) is immediate.

i).  (2.1b). The action of $\gc[ L,\cdot]$  on the lhs of (2.1b) is
 $$
 \eqalign{
 & [x,J([y,z])] + [J(x),[y,z]]+ ({\rm cyclic\  on\ } x,y,z )\cr
=& 2J([x,[y,z]]) + ({\rm cyclic\  on\ } x,y,z) =0 .\cr
}
\eqno(\rA.\num)
$$
On the other hand, the action of $\gc[ L,\cdot]$  on the rhs of (2.1b) is
zero because there is no $J(x)$ in it.

ii). (2.1c). Using (2.1a) and (2.1b), we have
$$
\eqalign{
&\gc \bigl[ L, [[J(x),J(y)],[z,J(w)]]  \bigr] \cr
=& [[x,J(y)],[z,J(w)]] + [[J(x),y],[z,J(w)]]
+ [[J(x),J(y)],[z,w]]
\cr
=&2  [J([x,y]),J(z')] -
[[J(y),z'],J(x)] - [[z',J(x)],J(y)]\qquad ({\rm where\ }z'=[z,w])
\cr
=& 3  [J([x,y]),J(z')]
+ \bigl\{ [ J(x),J([y,z'])] + [ J(y),J([z',x])] + [ J(z'),J([x,y])] \bigr\}
\cr
=& 3  [J([x,y]),J([z,w])]
+ \sum_{p,q,r} ([x,I_p],[[y,I_q],[[z,w],I_r]])
\{ I_p, I_q, I_r\}.
\cr
}
\eqno(\rA.\num)
$$
Thus the action of $\gc[ L,\cdot]$  on the lhs of (2.1c) is
$$
\sum_{p,q,r}\bigl( ([x,I_p],[[y,I_q],[[z,w],I_r]])
	+([z,I_p],[[w,I_q],[[x,y],I_r]])\bigr)\{ I_p,I_q,I_r \},
\eqno(\rA.\num)
$$
which is equal to the action of $\gc[ L,\cdot]$  on the rhs of (2.1c).

These calculations are actually a part of a proof of the fact
that  (4.4) is an automorphism of
$Y(g)$.
\reference
\bye